\documentclass[12pt]{article}
\usepackage{a4wide}
\usepackage{amstext,amsfonts,amssymb,amsmath}

\begin{document}
\begin{flushright}
FTUV--05-0125  \quad IFIC--05--10\\
January 25, 2005
\end{flushright}

\vspace{.5cm}

\begin{center}

\renewcommand{\thefootnote}{\fnsymbol{footnote}}

{\Large {\bf Superbranes, $D=11$ CJS supergravity and enlarged
superspace coordinates/fields correspondence}}\footnote{Talk
delivered at the {\it XIXth Max Born Symposium} on {\it
Fundamental Interactions and Twistor-like Methods}, Wroclaw, 28th
September--1st October, 2004, and at the Coral Gables Conference
{\it Miami 2004: celebrating 40 years of quarks, cosmology, CP
violation and physics conferences in greater Miami}, 15th
December-19th December, 2004. To appear in the Proc. of the Max
Born Symposium (AIP Proc. Series) and in the electronic Proc. of
the Coral Gables Conference.}

\vspace{1cm}

{\bf\large Jos\'e A. de Azc\'arraga}

\vspace{.5cm}

{\it Dept. of Theoretical Physics and IFIC (CSIC-UVEG),\\
           Facultad de F\'isica, 46100-Burjassot (Valencia), Spain}
\vspace{1cm}

\end{center}

\begin{abstract}
 We discuss the r\^ole of enlarged superspaces in two seemingly different
 contexts, the structure of the $p$-brane actions and that of the
 Cremmer-Julia-Scherk eleven-dimensional supergravity. Both provide examples
 of a common principle: the existence of an
 {\it enlarged superspaces coordinates/fields correspondence} by which all the
 (worldvolume or spacetime) fields of the theory are associated to
 coordinates of enlarged superspaces. In the context of $p$-branes, enlarged
 superspaces may be used to construct manifestly supersymmetry-invariant
 Wess-Zumino terms and as a way of expressing the Born-Infeld
 worldvolume fields of D-branes and the worldvolume M5-brane two-form
 in terms of fields associated to the coordinates of these enlarged
 superspaces. This is tantamount to saying that the Born-Infeld fields
 have a superspace origin,
 as do the other worldvolume fields, and that they have a composite
 structure. In $D$=11 supergravity theory enlarged superspaces
 arise when its underlying gauge structure is investigated and, as a
 result, the composite nature of the $A_3$ field is revealed:
 there is a full one-parametric family of enlarged superspace groups
 that solve the problem of expressing $A_3$ in terms of spacetime fields
 associated to their coordinates. The corresponding enlarged supersymmetry
 algebras turn out to be deformations of an {\it expansion} of the
 $osp(1|32)$ algebra. The unifying mathematical structure
 underlying all these facts is the cohomology of the
 supersymmetry algebras involved.
\end{abstract}


\renewcommand\thefootnote{\arabic{footnote}}
\setcounter{footnote}0

\newpage

\section{Introduction}

M-theory (see \cite{M-theory,M-alg} and \cite{Duff-hist} for a
chronological history) is not based at present on a definite
Lagrangian or on an S-matrix description; rather, its conjectured
existence relies on the properties of its six perturbative and low
energy limits (string models and supergravities) and by dualities
\cite{Duality} among them. Such dualities, including those
relating apparently different models, are believed to be
symmetries of M-theory. The full set of M-theory symmetries -as
the full M-theory itself- is not known\footnote{Several groups may
play a r\^ole, as the rank 11 Kac-Moody $E_{11}$ group
\cite{WestE11} as a basis for a non-linear realization approach to
$D=11$ supergravity as well as other Kac-Moody symmetries, or the
$OSp(1|64)$ group \cite{Bars,West00} and its subgroup $GL(32)$
\cite{BWest00,GGT}. This last one is the automorphism group of the
M-algebra $\{Q_\alpha , Q_\beta \} = P_{\alpha \beta}$; it is also
a manifest symmetry of the actions \cite{BL98,BPS04} for BPS
preons \cite{BPS01} (see \cite{Ba-05} for a review), the
hypothetical constituents of M-theory. Clearly, in $D=11$
supergravity one may see only a fraction of the M-theory
symmetries. As it was noticed recently \cite{DL03,Hull03} (see
also \cite{BPS04}), a suggestive analysis of supersymmetric $D=11$
supergravity solutions can be carried out in terms of generalized
connections with holonomy group $SL(32)$ (see \cite{Du-St} for an
early reference on generalized holonomy). The case for a
$OSp(1|32)\otimes OSp(1|32)$ gauge symmetry in a Chern-Simons
context was presented and discussed in
\cite{Zanelli,Horava,Nastase}.},
 but it should include these
dualities as well as the symmetries of the different superstring
and supergravity limits. For this reason, the study of the
symmetries of p-branes as well as the underlying gauge symmetry of
$D=11$ supergravity may help to understand the symmetry structure
of M-theory itself.

   Superalgebras going beyond the standard supersymmetry algebra
were considered very early (see \cite{vHvP82, D'A+F} and
references in \cite{A-I-bra-00}) and, later, in the context of
brane theory. Some of these {\it enlarged supersymmetry algebras}
generalize Green's algebra \cite{Green89}. They were introduced in
\cite{BeSez95} to make {\it Lie} algebras out of the free
differential algebras that had been introduced in \cite{AzTo89} to
recover cohomologically the classification \cite{AETW87} of the
scalar $p$-branes. The authors of \cite{BeSez95} also showed that
these algebras could be used to obtain Wess-Zumino (WZ) terms for
the rigid $p$-brane actions strictly invariant under
supersymmetry. The relation between semi- or quasi-invariance
({\it i.e.}, invariance but for a total derivative) of
lagrangians, cohomology and group extension theory, is a problem
which has a fifty years long history, but we will not discuss it
here (see \cite{CUP} and references therein). In the case of
$p$-branes, the additional variables of the new supersymetry
groups \cite{A-I-bra-00} (rigid enlarged superspaces) appear in
these manifestly invariant WZ terms in a trivial way, but this is
not the case for all types of branes. It was shown in
\cite{A-I-bra-00} that the enlarged superspaces, could also be
used to obtain Born-Infeld (BI) fields from one-forms defined on
them (for BI fields in the IIB case see \cite{Saka98}). In the
case of the D-branes BI fields or with the worldvolume two-form of
the M5 brane these fields allow for the existence of actions where
all worldvolume fields entering in the theory are associated to
the enlarged superspace coordinates (there are no fields
`external' to the superspace coordinates, {\it i.e.} {\it
directly} defined on the worldvolume). This points out to the
existence of an {\it extended superspace coordinates/fields
correspondence} for branes \cite{A-I-bra-00}, where in this case
`fields' refers to worldvolume fields. We shall devote the first
two sections to review these ideas.

  It turns out that, in an analogous fashion, a similar correspondence
may also be established for $D=11$ Cremmer-Julia-Scherk (CJS)
\cite{CJS} {\it supergravity}, in which case the fields are
spacetime fields. This correspondence between enlarged superspace
coordinates and spacetime fields for CJS supergravity is related
to the problem of its `hidden' or underlying gauge symmetry. This
problem was raised already in the original CJS paper \cite{CJS}.
It was considered by D'Auria and Fr\'e \cite{D'A+F} as a search
for a composite structure of the {\it three--form} $A_3$ field
that enters in the $D=11$ supergravity multiplet (see also
\cite{BAMcDo83,Kallosh84} for other discussions of the geometry of
$D=11$ supergravity and \cite{DuffEri04} for an overview on local
supersymmetry\footnote{For recent discussions in a different
perspective see \cite{Ma-Sa-03,Di-Fr-Mo-03,Le-Ma-03}.}). While the
graviton, the gravitino and the spin connection are one-form
fields, $e^a=dx^\mu e_\mu^a(x)$, $\psi^\alpha= dx^\mu
\psi^\alpha_\mu(x)$ and $\omega^{ab}= dx^\mu\omega_\mu^{ab}(x)$,
and can be considered as gauge fields for the superPoincar\'e
group \cite{DoMa77}, the fully antisymmetric
$A_{\mu_1\mu_2\mu_3}(x)$ (transverse) abelian gauge field is not
associated with a symmetry generator and it rather corresponds to
a {\it three}-form $A_3$ on spacetime. This prevents the
association of the fields of the standard $D=11$ supergravity
multiplet with the gauge fields of a Lie superalgebra, since these
are associated to {\it one}-forms.

   Two enlarged supersymmetry algebras with $528$ bosonic and $64$
fermionic generators
\begin{equation} \label{generators}
 \, P_a  \, , \, Q_\alpha  \; ; \;  Z_{a_1a_2}  \, ,
\, Z_{a_1 \ldots a_5}  \; ; \; Q^\prime_\alpha \; ,
\end{equation}
including the 528+32 M-algebra \cite{M-alg} ones plus a central
fermionic generator $Q^\prime_\alpha$, were found in \cite{D'A+F}
to allow for a decomposition of $A_3$. The corresponding one-form
fields
\begin{equation} \label{generators1}
 \, e^a  \, , \, \psi^\alpha  \; ; \;  B^{a_1a_2}  \, ,
\, B^{a_1 \ldots a_5}  \; , \; \eta^\alpha \; ,
\end{equation}
were then considered as gauge fields for these larger supergroups.
In this scheme, all the CJS supergravity fields can then be
treated as gauge fields, with $A_3$ expressed in terms of them.

   As we shall see, the problem studied in \cite{D'A+F} is
mathematically analogous to that of obtaining strictly invariant
WZ terms for $p$-branes from the originally quasi-invariant
(invariant up to a total derivative) ones. Both reduce to finding
a Lie algebra allowing us to write a closed invariant form (on the
original supergroup manifold) as the differential of an invariant
one (now on the manifold of the associated enlarged supergroup).
Expressed in another way, these problems correspond to finding a
trivialization of certain non-trivial Chevalley-Eilenberg (CE)
\cite{CE} cocycles for the cohomology of the standard
supersymmetry algebra of the theory by means of enlarging it. It
turns out that the underlying gauge supergroup structure of $D=11$
CJS supergravity can be described by any representative of a {\it
one--parametric} family of supergroups denoted $\tilde{\Sigma}(s)$
($s\not=0$), or by their associated superalgebras
$\tilde{\mathfrak{E}}(s)$, the two D'Auria-Fr\'e ones being two
particular elements of that family (specifically,
$\tilde{\mathfrak{E}}(3/2)$ and $\tilde{\mathfrak{E}}(-1)$). There
have been attempts to relate these solutions to some known algebra
(see \cite{CFGPvN}). We will see that the algebras
$\tilde{\mathfrak{E}}(s)$ are nontrivial ($s\not=0$) deformations
of the special element $\tilde{\mathfrak{E}}(0)$. The
$\tilde{\mathfrak{E}}(s), s\neq 0$, automorphism group is
$SO(1,10)$. Thus, the relevant supergroup that replaces the
standard superPoincar\'e group $\Sigma{\times\!\!\!\!\!\!\supset}
SO(1,10)$ becomes the semidirect product $\tilde{\Sigma}(s)
{\times\!\!\!\!\!\!\supset} SO(1,10)$, a deformation of
$\tilde{\Sigma}(0) {\times\!\!\!\!\!\!\supset} SO(1,10)$. As for
the superalgebra $\tilde{\mathfrak{E}}(0){+\!\!\!\!\!\!\supset}
so(1,10)$ itself, it is related to $osp(1|32)$ through an {\it
expansion}\footnote{The {\it expansion method} allows us to obtain
new algebras from a given one, in general of higher dimension than
the original one. Under a different name, expansions were
considered in \cite{H01}, and the method was studied in general in
\cite{AIPV02}.}. Specifically,
\begin{equation}
\tilde{\Sigma}(0) {\times\!\!\!\!\!\!\supset} SO(1,10) \approx
OSp(1|32)(2,3,2) \;  ,
\end{equation}
where the numbers in $OSp(1|32)(2,3,2)$ characterize the expansion
(see later). For $s = 0$ the $SO(1,10)$ automorphism group is
enhanced to $Sp(32)$, and one finds that $\tilde{\Sigma}(0)
{\times\!\!\!\!\!\!\supset} Sp(32) \approx OSp(1|32)(2,3)$.

The supergroup manifolds $\tilde{\Sigma}(s)$ determine {\it
rigid}, {\it enlarged} superspaces. The fact that all the
spacetime fields in (\ref{generators1}) may be associated to the
various coordinates of the $\tilde{\Sigma}(s)$ supergroups again
suggests that there is an {\it extended superspace
coordinates/fields correspondence} principle\footnote{The idea of
a `fundamental symmetry between coordinates and fields' is
explicitly stated in Berezin \cite{Be-79} and is implicit in
earlier work of D. V. Volkov \cite{Volkov-73}; the field space
democracy is also discussed in \cite{Sch-et-al-80}. However, we
are not referring here to a democracy between the fields and {\it
their} arguments (as one might introduce by {\it e.g.},
considering on an equal footing the coordinates of the {\it total}
space of a fibre bundle the cross sections of which may be used to
define fields on the {\it base} manifold), but rather to a
correspondence between the (enlarged) superspace coordinates and
the fields they originate, be them worldvolume or spacetime ones.
This is why is more precise to speak of a {\it correspondence}
between (enlarged superspace) coordinates and fields rather than
of `democracy' -the term used in \cite{A-I-bra-00}- since its
original use referred to a democracy between the fields and {\it
its} arguments. We have conjectured \cite{A-I-bra-00} the
existence of a correspondence between the {\it coordinates} of a
suitable superspace and the {\it fields} in theory constructed on
it. These appear as the pullbacks of forms, originally defined on
the target enlarged superspace, to the worldvolume or spacetime
manifolds. Also, the basis for such a correspondence is group
theoretical: the enlarged rigid superspaces are all supergroup
manifolds. }.


     \section{Wess-Zumino terms for super-$p$-branes and
           enlarged supersymmetry algebras}

In this section, and in the next one, we describe the r\^ole of
enlarged superspaces in brane theory. We will start from the
scalar $p$-branes \cite{AETW87} case. The action for a $p$ brane
in rigid superspace is given by the sum of two terms,
\begin{equation}
I=I_0+I_{WZ} \label{wz1} \;,
\end{equation}
where $I_0$ is the kinetic part and $I_{WZ}=\int_{\cal W}
\phi^*(b)$ is the WZ term, which is given by the integral over the
worldvolume ${\cal W}$, parametrized by $\xi^i=(\tau,
\sigma^1,\dots,\sigma^p)$ [$i=0,\dots,p$] of the pull-back
$\phi^*(b)$ to ${\cal W}$ of a $(p+1)$-form $b$ defined on the
rigid superspace $\Sigma^{(D|n)}$ ($\Sigma$ for short) of the
theory (the manifold of the corresponding supersymmetry group).
This form $b$ is the potential form of a ($p+2)$-form $h$ which
happens to be exact,
\begin{equation}
 h=db \label{wz2} \; ,
\end{equation}
and that is invariant under the transformations of the
superPoincar\'e group $\Sigma{\times\!\!\!\!\!\!\supset}
SO(1,D-1)$. The study of the different possible $(p+2)$-forms $h$
in superspaces corresponding to the minimal supersymmetries in $D$
dimensions determines the ($D,p$) values for which the WZ term
exists. With the appropriate relative factor, $I_{WZ}$ in
(\ref{wz1}) leads to super-$p$-brane $\kappa$-invariant actions.
These ($D,p$) values determine the `old branescan' \cite{AETW87}.

It  turns out \cite{AzTo89} that the $h$'s are non-trivial
Chevalley-Eilenberg (CE) $(p+2)$-cocycles  for the cohomology of
the standard ${\mathfrak{E}}^{(D|n)}$ supersymmetry algebra. This
means that $h$ is a closed (obviously, $dh=0$) and supersymmetry
invariant ($p+2$)-form constructed from the Maurer-Cartan (MC)
one-forms on the graded translations (supersymmetry) group
$\Sigma$ (namely $\Pi^\mu$ and $\Pi^\alpha$, where $\mu= 0,\dots
D-1$, and the range $1,\dots n$ of $\alpha$ depends on the minimal
spinor considered). This CE cocycle condition depends on the known
gamma matrix identities that are true only for the ($D,p$) values
of the `old branescan'. Furthermore, the non-triviality of these
cocycles means that the potential ($p+1$)-form $b$ in $h=db$ is
not supersymmetry invariant {\it i.e.}, that cannot be constructed
from the invariant MC forms $\Pi^\mu$ and $\Pi^\alpha$ on
$\Sigma$. An important consequence of this fact is that the WZ
lagrangian is not manifestly invariant under supersymmetry, but
only quasi-invariant (hence its `WZ' name) and that, as a result,
the algebra of charge densities produces topological extensions of
the original supersymmetry algebra \cite{AGIT89}.

Certain enlarged rigid superspaces associated to enlarged
supersymmetry groups ${\tilde \Sigma}$, with additional bosonic
and fermionic variables, can be used to obtain $p$-brane actions
that are equivalent to the standard ones but with WZ terms that
are strictly invariant under supersymmetry. On the manifolds of
these groups, the same $(p+2)$-forms $h$ of the old branescan are
still CE cocycles, but they are now trivial ones: $h=d{\tilde b}$,
and the new potential ($p+1$)-forms ${\tilde b}$ are ${\tilde
\Sigma}$-invariant. The process of obtaining these enlarged
algebras or `brane algebras' may be thus called of `trivialization
of the CE cocycles' $h$ on $\mathfrak{E}^{(D|n)}$.

Let us look in more detail how to achieve this trivialization (the
following is not, as we shall see, the  only possibility). One
starts with the MC equations of the standard supersymmetry algebra
in $D$ dimensions $\mathfrak{E}^{(D|n)}$ (we consider, for
simplicity, the cases that allow for real spinors; wedge products
are understood in this and in the next section)
\begin{equation}
         d\Pi^\alpha=0 \ , \quad d\Pi^\mu= a_s \Pi^\alpha
         (C\Gamma^\mu)_{\alpha\beta}\Pi^\beta \ ,
\label{wz3}
\end{equation}
where \footnote{{\it A comment on conventions}. The constant $a_s$
is real since we assume real gamma matrices (which in $D=11$, for
instance, requires mostly plus metric) and the convention used
here for the complex conjugation is $(\theta_1\theta_2)^* =
(\theta^*_1\theta^*_2)$, $\theta_1$ and $\theta_2$ being Grassmann
odd. If we used the conjugation that reverses the order, as it
will be the case in Secs. 4-7, the $a_s$ in eq. (\ref{wz3}) would
be purely imaginary. Other differences in conventions between
Secs. 2,3 and Secs. 4-7 are that in Secs.2,3 we write explicitly
the charge conjugation matrix $C$; also, in Secs. 2,3 the $\wedge$
product for forms is implicit. We have kept these two sets of
conventions in order to make direct contact with \cite{A-I-bra-00}
and with \cite{CJS-D=11}.} the constant $a_s=1/2$ for
$\tilde{\mathfrak{E}}$. The $(p+2)$-form $h$ for a $p$-brane may
be shown to be, up to a proportionality constant which is not
important in the discussion below,
\begin{equation}
   h=  \Pi^\alpha (C\Gamma_{\mu_1\dots\mu_p})_{\alpha\beta}
   \Pi^\beta \Pi^{\mu_1}  \dots \Pi^{\mu_p}
   \label{wz4}
\end{equation}
which is closed for the dimensions $D$ for which the identity
\begin{equation}
   (C\Gamma^{\mu_1\dots\mu_p})_{(\alpha\beta}
   (C\Gamma_{\mu_1})_{\gamma\delta)} =0
\label{wz5}
\end{equation}
is satisfied. The bilinear in (\ref{wz4}) suggests that in order
to find an invariant potential form for $h$ one should first
extend  (\ref{wz3}) adding the form $\Pi^{\mu_1\dots\mu_p}$ and
the MC equation
\begin{equation}
         d\Pi_{\mu_1\dots\mu_p} = a_0 \Pi^\alpha
(C\Gamma_{\mu_1\dots\mu_p})_{\alpha\beta}\Pi^\beta \ , \label{wz6}
\end{equation}
so that $h$ is the first term in the differential of
\begin{equation}
   \Pi_{\mu_1\dots\mu_p} \Pi^{\mu_1}  \dots \Pi^{\mu_p} \quad.
\label{wz7}
\end{equation}
Before going to the next step, let us note that this is a sensible
thing to do for two reasons. The first is that eq. (\ref{wz6}) and
the second of eqs. (\ref{wz3}) can be put on the same footing
since they are central (if one ignores the Lorentz part)
extensions of the abelian odd translation algebra defined by the
simple MC equations $d\Pi^\alpha=0$. So if the graded
supertranslations (supersymmetry) algebra is itself a central
extension, it seems mathematically natural to consider other
possible extensions as well. In fact, one should consider the most
general extension where the `central' generators appear for each
symmetric $(C\Gamma^{\mu_1\dots\mu_q})$ matrix (we shall keep,
here, however, only the generators corresponding to $\Pi^\mu$ and
one of the $\Pi^{\mu_1\dots\mu_p} $ for simplicity; they will be
sufficient to discuss the `scalar' brane actions). The second
reason is that including these new bosonic generators is necessary
to understand, from the algebraic point of view, the existence of
BPS states that break some supersymmetries but not all, as known
from supergravity theories.

Equation (\ref{wz7}) does not solve yet the problem of finding an
invariant $\tilde{b}$ such that $d \tilde{b}=h$ because the
exterior differential also acts on the $p$ factors $\Pi^\mu$ .
This means that new generators and MC equations have to be added
to (\ref{wz3}) and (\ref{wz7}).  It may be shown (see
\cite{BeSez95} and \cite{A-I-bra-00}) that such a $\tilde{b}$ can
be found if the algebra is extended in several steps,  each step
involving a central (if we ignore the Lorentz part) extension of
the algebra resulting from the previous one. The first invariant
form by which one extends is fermionic, and has the structure
$\Pi_{\mu_1 \dots \mu_{p-1}\alpha_1}$ (so that, for $p=1$, one
obtains the Green algebra \cite{Green89}). The second is an
extension of the algebra whose MC equations are generated by
$\Pi^\alpha$, $\Pi^\mu$, $\Pi_{\mu_1\dots\mu_p} $ and $\Pi_{\mu_1
\dots \mu_{p-1}\alpha_1}$, and the new invariant forms have the
structure $\Pi_{\mu_1 \dots \mu_{p-2}\alpha_1\alpha_2}$. This
process of  extensions ends when the last invariant form
$\Pi_{\alpha_1\dots \alpha_p}$ is added. At each step in the above
procedure the extension made is central, but it makes non-central
the former central generator of the previous step. Thus, the
resulting algebra is not a central extension of the supersymmetry
one but for $p=1$ where the only step produces the Green algebra
\cite{Green89}. The existence of these enlarged supersymmetry or
`brane algebras' depends on the values of $D$ and $p$ for which
the identity (\ref{wz5}) holds; this is not surprising since these
algebras allow for the existence of $\tilde{b}$ such that $d
\tilde{b}=h$ and $dh=0$ is true only when (\ref{wz5}) is
satisfied.

   We shall not give here the explicit expressions for the resulting
`brane algebras' in general; these can be found in \cite{BeSez95}
and in \cite{A-I-bra-00}; the associated enlarged superspace
groups law is also given in \cite{A-I-bra-00} for the most
interesting ones. We shall only write explicitly two of these
enlarged supersymmetry algebras, because they will be relevant in
the next sections. Let us begin by the superalgebra for $p=2$,
$D=11$ that trivializes the CE cocycle that defines the WZ term of
the $D=11$ membrane \cite{B+S+T1987}. It is given by the MC
equations
\begin{eqnarray}
      d  \Pi^\alpha &=& 0 \ ,\quad d\Pi^\mu =  \frac{1}{2}
      (C\Gamma^\mu)_{\alpha\beta}
\Pi^\alpha \Pi^\beta\ , \nonumber\\
 d\Pi^{\mu\nu} &=&  \frac{1}{2}
 (C\Gamma^{\mu\nu})_{\alpha\beta}\Pi^\alpha\Pi^\beta
 \ , \nonumber\\
d\Pi_{\mu\alpha} &=&  (C\Gamma_{\nu\mu})_{\alpha\beta}  \Pi^\nu
\Pi^\beta +
  (C\Gamma^\nu)_{\alpha\beta} \Pi_{\nu\mu} \Pi^\beta \ , \nonumber\\
d\Pi_{\alpha\beta} &=& -\frac{1}{2}
(C\Gamma_{\mu\nu})_{\alpha\beta} \Pi^\mu \Pi^\nu  -\frac{1}{2}
(C\Gamma^\mu)_{\alpha\beta} \Pi_{\mu\nu}\Pi^\nu \nonumber\\  & &
+\frac{1}{4}  (C\Gamma^\mu)_{\alpha\beta}
\Pi_{\mu\delta}\Pi^\delta  + (C\Gamma^\mu)_{\delta\alpha}
\Pi_{\mu\beta}\Pi^\delta  + (C\Gamma^\mu)_{\delta\beta}
\Pi_{\mu\alpha}\Pi^\delta \ . \label{wz8}
\end{eqnarray}
They can be obtained from the general expressions of
\cite{A-I-bra-00}, particularized to the case $p=2$, by suitably
fixing the undetermined constants. The above equations allow for the
existence of an invariant $\tilde{b}$ such that $d\tilde{b} = h =
\Pi^\alpha (C\Gamma^{\mu\nu})_{\alpha\beta} \Pi^\beta \Pi_\mu
\Pi_\nu$. The expression for $\tilde{b}$ is
\cite{BeSez95,A-I-bra-00}
\begin{equation}
     \tilde{b} = \frac{2}{3} \Pi_{\mu\nu} \Pi^\mu \Pi^\nu -
     \frac{3}{5} \Pi_{\mu\alpha} \Pi^\mu
  \Pi^\alpha -\frac{2}{15} \Pi_{\alpha\beta} \Pi^\alpha \Pi^\beta \ .
\label{wz9}
\end{equation}

The second algebra that we shall need is the one that trivializes
the CE cocycle associated to the WZ term of the D=10, IIA
superstring. It can be extracted from the dimensional reduction to
$D=10$ of the algebra (\ref{wz8}) (see \cite{A-I-bra-00}, eqs. (88)
and (90)), and is given by
\begin{eqnarray}
   d\Pi^\alpha&=&0 \ , \quad d\Pi^\mu= \frac{1}{2}
   (C\Gamma^\mu)_{\alpha\beta}
  \Pi^\alpha\Pi^\beta \nonumber\\
  d\Pi_\mu^{(z)} &=&  \frac{1}{2}
  (C\Gamma^\mu \Gamma_{11})_{\alpha\beta}
  \Pi^\alpha\Pi^\beta \nonumber\\
 d\Pi_\alpha^{(z)} &=& (C\Gamma_\nu\Gamma_{11})_{\alpha\beta}
 \Pi^\nu \Pi^\beta
+ (C\Gamma^\nu)_{\alpha\beta} \Pi_\nu^{(z)} \Pi^\beta \ ,
\label{wz10}
\end{eqnarray}
where the superscript $(z)$ refers to the fact that the forms
$\Pi_\mu^{(z)}$ and $\Pi_\alpha^{(z)}$ come from the dimensional
reduction of $\Pi_{\mu\nu}$ and $\Pi_{\mu\alpha}$ respectively
when $\mu$ corresponds to the $z$ coordinate in the splitting
$x^0,\dots,x^9,z$. This algebra is consistent due to the $D=10$
gamma matrices identity
\begin{equation}
      (C\Gamma^\mu \Gamma_{11})_{(\alpha\beta}
      (C\Gamma_\mu)_{\gamma\delta)} = 0 \; .
\label{wz11}
\end{equation}
 The corresponding $\tilde{b}$ is given by
\begin{eqnarray}
     \tilde{b} &=& \frac{1}{2} \Pi^\alpha\Pi_\alpha^{(z)} -
     \Pi^\mu\Pi_\mu^{(z)}\ ,\nonumber\\
  d \tilde{b} &=& h = (C\Gamma_\mu \Gamma_{11})_{\alpha\beta}
  \Pi^\mu \Pi^\alpha\Pi^\beta  \quad .
\label{wz12}
\end{eqnarray}

We note finally that the coordinates of ${\tilde \Sigma}/\Sigma$
[($\varphi^{\mu\nu}$,$\varphi_{\mu\alpha}$,$\varphi_{\alpha\beta}$)
for  eq. (\ref{wz8}) and ($\varphi^\mu$, $\varphi_\alpha$) for eq.
(\ref{wz10})]  that, beyond the ordinary superspace
$\Sigma^{(D|n)}$ ones ($x^\mu,\theta^\alpha$), complete the
parametrization the enlarged superspaces ${\tilde{\Sigma}}$, lead
to non-dynamical fields in the action. The WZ term is written in
invariant form as $\phi^* (\tilde{b})$, where $\phi^*$ is the
pullback that takes the form ${\tilde b}$ on the
${\tilde{\Sigma}}$ manifold to ${\cal W}$. Indeed, since $d{\tilde
b}=h=db$, it follows that the new fields enter in the WZ part of
the action
 $\int_{\cal W}\phi^*({\tilde b})$ as total derivative\footnote{We shall
not discuss the behaviour of the additional variables under
$\kappa$-symmetry, for which we refer to \cite{AIM-04}.}. However,
they appear non trivially in the context of D-branes and in the
M5-brane, as we discuss in the next section, where they also appear
in the $D$-brane action kinetic part.


\section{ The enlarged superspace coordinates/fields correspondence for
superbranes}

The action for D-branes
\cite{Ceder-et-al-96,Ag-Po-Sch-96,Berg-Town-97}
 (we shall restrict ourselves to the type IIA D-branes as in
 \cite{A-I-bra-00}, see
\cite{Saka98} for the IIB case) in a rigid background for which
all forms in the R-R sector and the dilaton vanish is given, as in
the case of $p$-branes (\ref{wz1}), by the sum of a kinetic term
$I_0$ and a WZ term $I_{WZ}$ . The first one is
\begin{equation}
I_0= \int d\xi^{p+1} \sqrt{-\det (g_{ij}+ \mathcal{F}_{ij})}
\quad.
\label{wz13}
\end{equation}
In (\ref{wz13}), $g_{ij}$ is the induced metric on the
worldvolume, $g_{ij}(\xi)=\Pi^\mu_i(\xi)\Pi_{\mu j}(\xi)$
($\Pi^\mu = \Pi^\mu_i d\xi^i$), and $\mathcal{F}_{ij}$ are the
worldvolume components of the form
$\mathcal{F}(\xi)=dA_1(\xi)-\phi^*(B_2)$, where $\phi^*(B_2)$ is
the pull-back to ${\cal W}$ of a two-form $(B_2)$ defined on the
$D=10$, IIA superspace such that
\begin{equation}
     dB_2=-(C\Gamma_\mu\Gamma_{11})_{\alpha\beta}\Pi^\alpha\Pi^\beta
     \Pi^\mu \ ,
\label{wz14}
\end{equation}
and $A_1(\xi)$ is a one-form {\it directly} defined on the
worldvolume, the BI field, that transforms in such a way that
$\mathcal{F}$ is invariant under supersymmetry. The WZ term is
quasi-invariant and is given by the integral of a $(p+1)$-form
that depends polynomially on $\mathcal{F}$, the coefficients being
forms on the $D=10$ IIA superspace. The explicit expressions for
the different D-brane WZ terms (actually for even $p=2,4,6,8$) are
not relevant for our purposes, but we note here that the search
for the possible non-trivial CE cocycles that determine $h$ also
identifies the possible D-branes \cite{A-I-bra-00} recovering
Polchinski's classification \cite{Pol-95} (for recent work on
D-branes see \cite{An-Gra-03+Ha-Ka-04}).
  Similarly, the $D=11$ M5-brane action \cite{BPSTV95}
\footnote{The covariant equations for the D-branes and the
M5-brane were found in \cite{Ho-Se-96} in the framework of the
superembedding approach developed for the supermembrane and
superstrings in \cite{Ba-So-To-Pa-Vo-95}.} contains a {\it
two}-form $A_2(\xi)$ directly defined on the worlvolume that
enters the action through the field strength $H_3(\xi)=dA_2(\xi)-
\phi^*({\cal A}_3)$, where $\phi^*({\cal A}_3)$ is the pull back
to the worldvolume of a $D=11$ superspace three-form ${\cal A}_3$
such that
\begin{equation}
        d{\cal A}_3=- (C\Gamma_{\mu\nu})_{\alpha\beta} \Pi^\mu\Pi^\nu
        \Pi^\alpha\Pi^\beta \; ;
\label{wz15}
\end{equation}
$A_2(\xi)$ transforms under supersymmetry in such a way that $H$
is invariant \footnote{Clearly, eq. (\ref{wz15}) shows that the
standard superspace three-form ${\cal A}_3$ cannot be invariant
under the transformations of standard supersymmetry. The ${\cal
A}_3$ here is, up to a factor, the $\omega_3$ in eq. (\ref{a4=0})
and corresponds to the $A_3$ form in the case of curved
susperspace.}.
 Again we note in passing that a CE cohomological search
for the possible $D=11$ WZ terms in this case leads to the M5
brane as the only solution \cite{A-I-bra-00}.

  In contrast with $p$-branes, both the D-brane and M5-brane
actions cannot be written in terms of forms associated to the
coordinates of ordinary superspaces, due to the presence of the
one- and two-forms $A(\xi)$, which are defined directly on the
worldvolume. The arguments of the previous section, however, lead
to the possibility of writing them solely in terms of forms
defined on suitably {\it enlarged} superspaces as we describe now.

   Let us first consider the IIA D-branes case (this includes the D2,
D4, D6 and D8 cases). The two-form $\mathcal{F}$ is supersymmetry
invariant and has the property that $\mathcal{F}= \phi^*(
(C\Gamma_\mu\Gamma_{11})_{\alpha\beta}\Pi^\alpha\Pi^\beta
\Pi^\mu)$. But these conditions are also satisfied by $\tilde{b}$
in eq. (\ref{wz12}). Moreover, as discussed in the previous
section, the new superspace variables appear in $\tilde{b}$ inside
a total differential. So one may consider the IIA  enlarged
superspace defined by the MC equations (\ref{wz10}) and identify
$\mathcal{F}$ with $\phi^*(\tilde{b})$. Since $\tilde{b}$ has a
part that is a total differential which contains the new
superspace coordinates, one concludes that $dA$ may be identified
with this part. The result \cite{A-I-bra-00} is
\begin{equation}
     A_1(\xi) = \phi^*\left(\varphi_\mu dx^\mu + \frac{1}{2}
     \varphi_\alpha d\theta^\alpha\right)\ .
     \label{wz16}
\end{equation}
In the M5-brane case, the relevant enlarged superspace is the one
corresponding to the MC equations (\ref{wz8}), and the expression
for the two-form $A_2(\xi)$ that arises by identifying $H$ with
$\phi^*(\tilde{b})$ from eq. (\ref{wz9}) turns out to be
\cite{A-I-bra-00}
\begin{eqnarray}
    A_2(\xi) &=& \phi^*\left(\frac{2}{3}\varphi_{\mu\nu} dx^\mu dx^\nu
    - \frac{3}{5} \varphi_{\mu\alpha} dx^\mu d\theta^\alpha -
    \frac{2}{15} \varphi_{\alpha\beta} d\theta^\alpha
    d\theta^\beta+ \frac{1}{30} \varphi_{\mu\nu} x^\mu
    (C\Gamma^\nu)_{\alpha\beta} d\theta^\alpha d\theta^\beta
    \right.
    \nonumber \\
    & & + \frac{11}{30} \varphi_{\mu\nu} dx^\mu
    (C\Gamma^\nu)_{\alpha\beta} \theta^\alpha d\theta^\beta -
    \frac{13}{180}\varphi_{\mu\nu} (C\Gamma^\mu)_{\alpha\beta}
    (C\Gamma^\nu)_{\delta\epsilon} \theta^\alpha d\theta^\beta
    \theta^\delta d\theta^\epsilon \nonumber\\
    & & \left. +\frac{1}{10} \varphi_{\mu\alpha}
    (C\Gamma^\mu)_{\delta\epsilon} \theta^\delta d\theta^\epsilon
    d\theta^\alpha + \frac{1}{20}
    \varphi_{\mu\alpha}(C\Gamma^\mu)_{\delta\epsilon}d\theta^\delta
    d\theta^\epsilon \theta^\alpha\right) \ .
\label{wz17}
\end{eqnarray}

So far we have argued that $\phi^*(\tilde{b})$ has the same
supersymmetry properties as $\mathcal{F}(\xi)$ (for the IIA
D-branes) or $H(\xi)$ (for the M5 brane). The next question to ask
is whether it is legitimate to substitute the latter for the
former in the actions for the D-branes and the M5-brane. Let us
begin by the D-brane case. It suffices to show that the
Euler-Lagrange (E-L) equations for the actions that are obtained
by replacing $A_1(\xi)$ by the r.h.s. of (\ref{wz16}), which we
shall denote more explicitly by
$A_1(x(\xi),\theta(\xi),\varphi(\xi))\equiv
A_1(x,\theta,\varphi)$, have the same dynamical content as those
for the original one, where $I=I[x,\theta,A_1]$. Indeed,
$I[x,\theta,A_1(x,\theta,\varphi)]$ has the same variation as
$I[x,\theta,A_1]$, except for the fact that one has to vary also
the fields inside $A_1(x,\theta,\varphi)$. From the variation of
$I[x,\theta,A_1(x,\theta,\varphi)]$ with respect to the new
variables, $\varphi=(\varphi_\mu,\varphi_\alpha)$, one arrives at
$\Pi^\mu_i \left.\frac{\delta I[A_1]}{\delta
A_{1i}(\xi)}\right\vert_{A_1=A_1(x,\theta,\varphi)}= 0$, and this
leads to $\left.\frac{\delta I[A_1]}{\delta
A_{1i}(\xi)}\right\vert_{A_1=A_1(x,\theta,\varphi)}=0$ provided
that the induced metric on the worldvolume is nondegenerate, as is
always the case in tensionful brane theory. But this last equation
is one of the equations for $I[x,\theta,A_1]$. If we now
substitute it in the equations for
$I[x,\theta,A_1(x,\theta,\varphi)]$, and use the fact that the
variations through $A_1(x,\theta,\varphi)$ are proportional to
$\left.\frac{\delta I[A_1]}{\delta
A_i(\xi)}\right\vert_{A_1=A_1(x,\theta,\varphi)}$ due to the chain
rule, we recover the remaining equations of $I[x,\theta,A_1]$.

    Moreover, since in $I[x,\theta,A_1(x,\theta,\varphi)]$ all the new
variables appear inside $A_1$, the extra degrees of freedom
corresponding to them have to be reduced by a gauge symmetry to
those of the customary BI field $A_{1i}(\xi)$. Since $\frac{\delta
I}{\delta \varphi_\alpha}=0$ is itself a Noether identity (which
follows by looking at the E-L equations as above), by the second
Noether theorem there is a gauge symmetry allowing us to remove
$\varphi_\alpha$ entirely. We also note that the $D$ equations
$\left.\frac{\delta I[A_1]}{\delta \varphi_\mu
}\right\vert_{A_1=A_1(x,\theta,\varphi)}$  produce only ($p+1$)
independent ones $\frac{\delta I[A]}{\delta A_{1i}(\xi)}=0$ and,
consequently, the remaining [$D-(p+1)$] equations are Noether
identities. Hence, of  the $D$ extra bosonic degrees of freedom
variables introduced by $\varphi_\mu$ in
$A_{1i}(x,\theta,\varphi)$, [$D-(p+1)$] may be eliminated by a
gauge transformation. To see this explicitly, let us set
$\varphi_\alpha=0$ and write $A_{1i}(x,\varphi)=
\varphi_\mu\partial_i x^\mu$ in a local gauge such that
$x^0=\tau,x^1=\sigma^1,\dots , x^p=\sigma^p$ and let $x^K(\xi)$ be
the remaining $x$'s. Then $A_i(\varphi,x,\theta)= \varphi_i(\xi) +
\varphi_K(\xi)
\partial_i x^K (\xi)$, $K=(p+1),\dots, D-1$.  Thus,
$A_{1i}(\varphi,x,\theta)$, and therefore the action, remains
invariant under the following set of $D-(p+1)$ gauge
transformations:
\begin{equation}
  \delta \varphi_K(\xi) = \alpha_K(\xi) \quad ,\quad
  \delta \varphi_i(\xi) = - \alpha_K(\xi)\partial_i x^K(\xi)
  \quad .
 \label{wz18}
\end{equation}
By taking $\alpha_K=\varphi_K$ we find that $\varphi_K=0$ so that
$A_i(\xi)=\varphi_i(\xi)$. Thus, the actual number of degrees of
freedom is ($p+1$), and no new dynamical ones are added by
assuming the compositeness $A_{1i}(x,\theta,\varphi)$ of the BI
fields. The case of the M5 brane can be treated similarly
\cite{A-I-bra-00}.

   We show next that an analogous mechanism is at work in D=11 supergravity
when it is expressed in terms of a composite three form $A_3$ by
using the coordinates of a suitably enlarged superspace.


\section{$D=11$ supergravity and composite nature of the $A_3$ field}

We turn now to $D=11$ supergravity, a different theory that
nevertheless presents several analogies with the previous
discussion on branes. Our aim here is to find a composite
structure of the $A_3$ field of CJS supergravity. This problem is
equivalent to that of trivializing a four-cocycle ($\omega_4$
below) for the standard supersymmetry algebra cohomology on a
larger superalgebra, so that $\omega_4=d{\tilde\omega}_3$ where
${\tilde\omega}_3$ is expressed in terms of MC one-forms on the
corresponding larger superspace group manifold. In this way $A_3$
will not be `external' to the superspace coordinates of the theory
and, at the same time, the enlarged superymmetry algebra will
reveal the hidden underlying gauge symmetry of CJS supergravity.

The field content of Cremmer-Julia-Scherk $D=11$ supergravity
multiplet is the (unique) $D=11$ supergravity one
\begin{equation}
\label{CJSmulti}
 ( \; e^a(x) \; , \; \psi^\alpha(x) \; , \; A_3(x) \;)
\end{equation}
where $e^a(x)$ is the {\it elfbein}, $\psi^\alpha(x)$ (a Majorana
spinor) is the gravitino field and $A_3(x)$ is an antisymmetric
three-index Abelian gauge field. The first order formulation of
$D=11$ supergravity further requires an initially independent spin
connection $\omega^{ab}(x)$.

   As is well known, a justification for this set of fields is provided
by the on-shell counting of bosonic and fermionic degrees of
freedom. By considering the transverse traceless spatial ($D=11$)
components of $g_{ij}$ and those of $A_{ijk}$, one finds that
$e^a$ has $\frac{(D-2)(D-1)}{2} - 1 = \frac{D(D-3)}{2} = 44$
bosonic d.o.f. and that $A_3$ has ${D-2 \choose 3} = 84$ bosonic
ones; as for $ \psi^\alpha$, it has $\frac{1}{2} 2^{[D/2]} (D-3) =
\frac{1}{2} 32 (11-3) = 128$ fermionic d.o.f.  Thus, as it should
be the case because of the supersymmetry of the theory, the
numbers of bosonic and fermionic d.o.f.~match:
\begin{eqnarray*} \sharp \; \textrm{Bosonic d.o.f.} \; = 44 + 84
=  128 = \; \sharp \; \textrm{Fermionic d.o.f.}\quad .
\end{eqnarray*}

The supergravity one-forms $e^a$, $\psi^\alpha$ and $\omega^{ab}$
generate a free differential algebra\footnote{In essence, a FDA
(introduced in this context in \cite{D'A+F} as a {\it Cartan
integrable system}) is an exterior algebra of forms, with constant
coefficients, that is closed under the exterior derivative $d$; see
\cite{Su77,D'A+F,Ni83,Cas-D'A-Fre-book}.} (FDA) . This is defined by
the expressions for the FDA curvatures
\begin{eqnarray}\label{Ta=}
{\mathbf R}^a &:=&  de^a -e^b\wedge \omega_b{}^a  + i\psi^{\alpha}
\wedge \psi^{\beta} \Gamma^a_{\alpha\beta}  \; , \\
\label{Tal=} {\mathbf R}^\alpha  &:= &  d\psi^\alpha - {1\over 4}
\psi^\beta \wedge  \omega^{ab} \Gamma_{ab\; \beta}{}^\alpha \; ,
\\
\label{Rab=} \mathbf{R}^{ab} &:= &  d \omega^{ab}  - \omega^{ac}
\wedge \omega_c{}^{b}  \; ,
\end{eqnarray}
where $T^a:=De^a=de^a -e^b\wedge \omega_b{}^a$ is the torsion and
$\mathbf{R}^{ab}$ coincides with the Riemann curvature, and by the
Bianchi identities (which are the consistency/integrability
conditions for the FDA).

For vanishing curvatures, ${\mathbf R}^a=0$, ${\mathbf R}^\alpha
=0$, ${\mathbf R}^{ab} =0$,  eqs.~(\ref{Ta=}) and (\ref{Tal=}),
(\ref{Rab=}) reduce to the MC equations for the superPoincar\'e
algebra. Removing the unessential Lorentz part one arrives to the
MC equations for the graded translations (supersymmetry) algebra
$\mathfrak{E}^{(11|32)}$ ($\mathfrak{E}^{(D|n)}$ in general) (see
footnote $5$ for conventions)
\begin{eqnarray}\label{MCTa=}
de^a = - i\psi^{\alpha} \wedge \psi^{\beta} \Gamma^a_{\alpha\beta}
\; , \qquad d\psi^\alpha =0 \; ,
\end{eqnarray}
which correspond to the supersymmetry algebra commutation relations,
\begin{eqnarray}
\label{QQ=P} {} \{ Q_\alpha , Q_\beta \} = \Gamma^a_{\alpha\beta}
P_a \; , \qquad [P_a, Q_\alpha]=0 \; , \quad [P_a, P_b]=0 \; .
\end{eqnarray}

Eq.~(\ref{MCTa=}) is solved by
\begin{eqnarray}\label{ea=Pia}
e^a = \Pi^a := dx^a - i d\theta^\alpha  \Gamma^a_{\alpha\beta}
\theta^\beta \; , \qquad \psi^{\alpha} = \Pi^{\alpha} :=
d\theta^\alpha \; .
\end{eqnarray}
When $\Pi^a$, $\Pi^\alpha$ are considered as forms on the rigid
superspace $\Sigma^{(11|32)}$ ($\Sigma^{(D|n)}$ in general)
parametrized by $Z^M=(x^a,\theta^\alpha)$, they define the invariant
MC forms of the supertranslation algebra (\ref{QQ=P}) on the {\it
standard supersymmetry group manifold} $\Sigma^{(11|32)}$ that may
be identified with rigid superspace. When $e^a$ and $\psi^\alpha$
are forms on spacetime, the $x^a$ are still spacetime coordinates
while the $\theta^\alpha$ are Grassmann functions,
$\theta^\alpha(x)$, the Volkov-Akulov Goldstone fermions
\cite{VA72}. For one-forms defined on the standard curved
superspace, $e^a=dZ^ME_M^a(Z)$, $\psi^\alpha= dZ^ME_M^\alpha(Z)$,
$\omega^{ab}(Z)= dZ^M\omega_M^{ab}(Z)$ the FDA (\ref{Ta=}),
(\ref{Tal=}), (\ref{Rab=}) with nonvanishing ${\mathbf R}^\alpha$
and ${\mathbf R}^{ab}= R^{ab}$ but vanishing ${\mathbf R}^a=0$ gives
a set of superspace supergravity constraints (which are kinematical
or {\it off-shell} for $N=1,D=4$, and {\it on-shell}, {\it i.e.}
containing equations of motion among their consequences, for higher
$D$ including $D=11$). Nevertheless, the FDA makes also sense for
forms on spacetime, where $e^a=dx^\mu e_\mu^a(x)$ and
$\psi^\alpha=dx^\mu \psi_\mu^\alpha(x)$ are the gauge fields for the
supertranslations group.

   However, the $D=11$ supermultiplet (\ref{CJSmulti}) also includes
the three-form $A_3$, and the previous FDA generated by the
one-forms $e^a$, $\psi^\alpha$ and $\omega^{ab}$ has to be
completed by the definition of the four--form field strength
\cite{D'A+F}
\begin{eqnarray}\label{R4=} \mathbf{R}_4 &=&
dA_3 + {1\over 4} \psi^\alpha \wedge \psi^\beta \wedge e^a \wedge
e^b \Gamma_{ab}{}_{\alpha\beta} \; .
\end{eqnarray}
Note that, considering the FDA (\ref{Ta=}), (\ref{Tal=}),
(\ref{Rab=}), (\ref{R4=}) on the $D=11$ superspace and setting
$\mathbf{R}^a=0$ and $\mathbf{R}_4=F_4:= 1/4! e^{a_4}\wedge \ldots
\wedge e^{a_1}F_{a_1\ldots a_4}$ one arrives at the original
on-shell $D=11$ superspace supergravity constraints
\cite{CremmerFerrara80,BrinkHowe80} (see also
\cite{Lechner93,Howe97}).

  Thus, in contrast with the $D=4$ case, the $D=11$ supergravity FDA for
vanishing curvatures cannot be associated with the MC {\it
one}-forms and equations of a {\it Lie} superalgebra due to the
presence of the {\it three}-form $A_3$. On $\Sigma^{(11|32)}$, where
one also sets $\mathbf{R}_4=0$ by consistency, $dA_3$ becomes the
bosonic four-form
\begin{eqnarray}\label{a4=}
a_4= - {1\over 4} \psi^\alpha \wedge \psi^\beta \wedge e^a \wedge
e^b \Gamma_{ab}{}_{\alpha\beta} \;,
\end{eqnarray}
which corresponds to CE Lie algebra cohomology {\it four-cocycle} on
the standard supersymmetry algebra $\mathfrak{E}^{(11|32)}$, {\it
i.e.}
\begin{equation}\label{a4=0}
\omega_4 = - {1\over 4} \Pi^\alpha \wedge \Pi^\beta \wedge \Pi^a
\wedge \Pi^b \Gamma_{ab}{}_{\alpha\beta}=d\omega_3(x,\theta) \equiv
-{1\over 4} d\theta^\alpha \wedge d\theta^\beta \wedge \Pi^a \wedge
\Pi^b \Gamma_{ab}{}_{\alpha\beta} \; .
\end{equation}
This is so because $\omega_4$ is 1) $\Sigma^{(11|32)}$--invariant
and 2) closed. The four-cocycle $\omega_4$ is, furthermore, CE
non-trivial, since $\omega_3$ cannot be expressed in terms of the
$\mathfrak{E}^{(11|32)}$ MC forms: $\omega_3$ is not
$\Sigma^{(11|32)}$-invariant. However, it will be seen that there
exists \cite{CJS-D=11} a one-parametric family of {\it extended}
superalgebras $\tilde{\mathfrak{E}}(s)$, with MC forms defined on
the associated extended superspace group $\tilde{\Sigma}(s)$
manifolds, on which the CE four-cocycle $\omega_4$ becomes trivial
{\it i.e}, $\omega_4=d\tilde{\omega}$ and $\tilde{\omega}$ is made
out of $\tilde{\Sigma}(s)$-invariant MC forms. Of course, we have
already mentioned another solution for the same mathematical
problem in the context of branes, eq. (\ref{wz9}), but here we
shall concentrate on $\tilde{\mathfrak{E}}(s)$ due to their closer
relation with $osp(1|32)$ and with the M-theory superalgebra
(itself an expansion, osp(1|32)(2,1,2), of $osp(1|32)$
\cite{AIPV02} ). Substituting in the expression of
${\tilde\omega}_3$ the gauge fields for the MC forms, the
resulting expression will provide the composite structure of the
$A_3$ form in terms of $\tilde{\mathfrak E}(s)$ gauge fields.

Thus, formulated in this way, the problem of writing the $A_3$
field in terms of one-form fields is, like the construction of WZ
invariant terms for branes or the search for an (enlarged)
superspace origin of the BI fields, purely geometrical: it reduces
to a problem of Lie superalgebra cohomology. It is equivalent, in
the spirit of the {\it enlarged superspace coordinates/fields
correspondence}, to looking for an enlarged supergroup manifold
$\tilde{\Sigma}$ on which one can find a suitable invariant
three-form $\tilde{\omega}_3$ (corresponding to $A_3$) written in
terms of products of $\tilde{\mathfrak{E}}$ MC forms on
$\tilde{\Sigma}$ (which will give rise to the one-form gauge
fields). The three-form $\tilde{\omega}_3$ will necessarily depend
on the coordinates of the generalized (extended) superspace group
manifold $\tilde{\Sigma}$; in contrast, the original $\omega_3
=\omega_3(x^a, \theta^\alpha)$ depends on the coordinates of
standard superspace $\Sigma^{(11|32)}$. The MC equations of the
enlarged superspace algebra $\tilde{\mathfrak{E}}$ can be
`softened' by adding the appropriate curvatures. The resulting
gauge FDA for the `soft' forms over $D=11$ spacetime will then
describe a $D=11$ supergravity theory in which $A_3$ is a {\it
composite}, not elementary, field.

To finish this section, we remark that, although the composite
nature of $A_3$ and the superspace origin of the worldvolume
fields in the D-branes and M5 brane depend on the same
mathematical problem as stated above, the relevant objects
involved in those two problems are different: whereas ${\tilde
\omega_3}$ above, being constructed entirely in terms of the MC
forms of certain enlarged supergroups is invariant under its
transformations, the {\it e.g.} BI fields expressed through the
coordinates of the enlarged superspaces are not invariant under
the corresponding enlarged supersymmetry (see eqs. (\ref{wz16})
and (\ref{wz17})); only the forms $\mathcal{F}$ and $H$ are
invariant.


\section{Trivialization of the CE four-cocycle}

Let us describe now the solution of the trivialization problem just
described. We shall first write down the algebras suitable to this
end, and then the expression for $\tilde{\omega}_3$, which also
gives the composite structure of $A_3$. At the end of this section,
we shall specialize the results for a particularly simple case.

\subsection{ A family of extended superalgebras $\tilde{\mathfrak
E}(s)$ }

The three-form $A_3$ of the $D=11$ supergravity FDA may be written
in terms of one-forms by introducing {\it two} new {\it bosonic}
tensorial one-forms, $B^{a_1 a_2}$, $B^{a_1 \ldots a_5}$, and
\emph{one} new {\it fermionic} spinorial one-form, $\eta^\alpha$,
that obey the FDA equations (\ref{Ta=})--(\ref{Rab=}), (\ref{R4=})
plus
\begin{eqnarray}
\label{cBab} {\cal B}_2^{a_1a_2} &=& DB^{a_1a_2} + \psi^\alpha
\wedge \psi^\beta \, \Gamma^{a_1a_2}_{\alpha\beta} \; , \qquad
\\
\label{cB15} {\cal B}_2^{a_1\ldots a_5} &=& DB^{a_1\ldots a_5} + i
\psi^\alpha  \wedge \psi^\beta \,
\Gamma^{a_1\ldots a_5}_{\alpha\beta} \; , \qquad \\
 \label{cBal}
{\cal B}_2^\alpha  &=& D\eta^{\alpha} - i \, \delta \; e^a \wedge
\psi^\beta \Gamma_{a\, \beta}{}^\alpha \nonumber \\ &-& \gamma_1
\, B^{ab} \wedge \psi^\beta \Gamma_{ab\, \beta}{}^\alpha - i \,
\gamma_2 \, B^{a_1\ldots a_5} \wedge \psi^\beta \Gamma_{a_1\ldots
a_5 \beta}{}^\alpha \; \qquad
\end{eqnarray}
where $\gamma_1$, $\gamma_2$ and $\delta$ are parameters (that are
related by eq.~(\ref{idg}) below).

 For vanishing curvatures (and ignoring the spin
connection) the above FDA reduces to the MC equations
\begin{eqnarray}
de^a &=& - i\psi^{\alpha} \wedge \psi^{\beta}
\Gamma^a_{\alpha\beta} \; , \quad d\psi^\alpha =0 \; , \label{MC1}\\
\label{dB2=} dB^{a_1a_2} &=& - \psi^\alpha \wedge \psi^\beta \,
\Gamma^{a_1a_2}_{\alpha\beta} \; , \quad  dB^{a_1\ldots a_5} = - i
\psi^\alpha \wedge
\psi^\beta \, \Gamma^{a_1\ldots a_5}_{\alpha\beta} \; ,  \\
 \label{deta=}
d\eta^{\alpha} &=& \psi^\beta \wedge  {\left(- i \, \delta \, e^a
\Gamma_{a}- \gamma_1 \, B^{ab} \Gamma_{ab} - i \, \gamma_2 \,
B^{a_1\ldots a_5} \Gamma_{a_1\ldots a_5} \right)_\beta}^\alpha \;
, \qquad
\end{eqnarray}
which correspond to the $D=11$ superalgebra commutators
\begin{eqnarray}\label{susyalg}
 & \{Q_\alpha,Q_\beta\}= P_{\alpha\beta} := \Gamma^a_{\alpha\beta}
P_a + i\Gamma^{a_1a_2}_{\alpha\beta} Z_{a_1a_2} +
\Gamma^{a_1\ldots a_5}_{\alpha\beta} Z_{a_1\ldots a_5} \; , \\
\label{[Z,Q]}
 & [ P_a , Q_\alpha ] = \delta \;  \Gamma_{a \; \alpha}{}^\beta
Q^\prime_\beta \; , \quad \nonumber
\\ & [ Z_{a_1a_2} , Q_\alpha ] = i\gamma_1
\Gamma_{a_1a_2 \; \alpha}{}^\beta Q^\prime_\beta \; , \; [ Z_{a_1
\ldots a_5} , Q_\alpha ] =\gamma_2 \Gamma_{a_1 \ldots a_5 \;
\alpha}{}^\beta Q^\prime_\beta \;   .
\end{eqnarray}
The constants $\delta$, $\gamma_1$,  $\gamma_2$ are clearly
restricted by the Jacobi identities, which require
\begin{equation}
\label{idg} \delta + 10 \gamma_1- 6! \gamma_2 = 0 \; .
\end{equation}
One non-vanishing parameter ({\it e.g.}, $\gamma_1$) can be
removed by rescaling the new fermionic generator $Q^\prime_\alpha$
and it is thus inessential. As a result,
eqs.~(\ref{susyalg})--(\ref{idg}) describe, effectively, a {\it
one-parameter family} of Lie superalgebras, denoted $\tilde
{\mathfrak E}(s)$. The parameter $s$ may be introduced, {\it e.g.}
through
\begin{eqnarray}
\label{s-def} s:= {\delta \over 2\gamma_1} - 1   \qquad &
\Rightarrow & \qquad \left\{
{\setlength\arraycolsep{2pt}\begin{array}{lll} \delta & =  & 2\gamma_1(s+1) \, , \\
\gamma_2 & = & 2\gamma_1({s \over 6!} + {1 \over 5!}) \; .
\end{array}} \right.    \;
\end{eqnarray}
In this parametrization the element corresponding to the case
$\gamma_1 = 0$ may be included as the $\gamma_1 \rightarrow 0$ limit
with $\gamma_1 s \rightarrow \delta/2 \neq 0$. This implies that the
corresponding algebra (labelled $\tilde{\mathfrak{E}}(\infty)$) is a
regular member of the family.

In terms of $s$, eqs.~(\ref{[Z,Q]}) read:
\begin{eqnarray}\label{Sigma(s)}
& [ P_a , Q_\alpha ]  =  2\gamma_1(s+1) \;  \Gamma_{a \;
\alpha}{}^\beta Q^\prime_\beta \; , \quad \nonumber
\\ & [ Z_{a_1a_2} , Q_\alpha ]  =  i\gamma_1
\Gamma_{a_1a_2 \; \alpha}{}^\beta Q^\prime_\beta \; , \nonumber
\\  & [ Z_{a_1 \ldots a_5} , Q_\alpha ]  =  2\gamma_1({s \over 6!} + {1
\over 5!}) \Gamma_{a_1 \ldots a_5 \; \alpha}{}^\beta
Q^\prime_\beta \; .
\end{eqnarray}
The family $\tilde {\mathfrak E}(s)$ is equivalently defined by
its MC equations {\setlength\arraycolsep{1pt}
\begin{eqnarray}\label{MCSigma(s)}
de^a &=& - i\psi^{\alpha} \wedge \psi^{\beta}
\Gamma^a_{\alpha\beta} \;, \quad d\psi^\alpha =0 \; , \; \nonumber\\
 dB^{a_1a_2} &=& - \psi^\alpha
\wedge \psi^\beta \,
\Gamma^{a_1a_2}_{\alpha\beta} \; , \qquad \nonumber\\
dB^{a_1\ldots a_5} &=& - i \psi^\alpha \wedge
\psi^\beta \, \Gamma^{a_1\ldots a_5}_{\alpha\beta} \; ,  \nonumber\\
d\eta^{\alpha} &=& -2\gamma_1\psi^\beta \wedge \left( i \, (s+1)
\, e^a \Gamma_{a\, \beta}{}^\alpha \right. \nonumber \\  & +&
\left. \frac12 \, B^{ab} \Gamma_{ab\, \beta}{}^\alpha + i \,
\left({s \over 6!} + {1 \over 5!}\right) \, B^{a_1\ldots a_5}
\Gamma_{a_1\ldots a_5 \beta}{}^\alpha \right)  . \quad
\end{eqnarray}}

   The $s=0$ is a special case. The $\tilde {\mathfrak E}(0)$
superalgebra is given by
\begin{eqnarray}
\label{osp-exp} {} \{Q_\alpha,Q_\beta\}= P_{\alpha\beta} \; ,
\quad [P_{\alpha\beta} , Q_{\gamma}] = 64 \; \gamma_1 \; C_{\gamma
(\alpha } Q^\prime_{\beta )} \; ,
\end{eqnarray}
which are obtained from eqs.~(\ref{Sigma(s)}) with $s=0$ by using
the Fierz identity
\begin{eqnarray*}
 \label{II=GG}
\delta_{(\alpha}{}^{\gamma} \delta_{\beta)}{}^{\delta} &=& {1\over
32} \left( \Gamma^a_{\alpha\beta} \Gamma_a^{\gamma\delta} -
{1\over 2}
\Gamma^{a_1a_2}{}_{\alpha\beta}\Gamma_{a_1a_2}{}^{\gamma\delta} +
{1\over 5!} \Gamma^{a_1\ldots a_5}{}_{\alpha\beta}
\Gamma_{a_1\ldots a_5}{}^{\gamma\delta}\right) .
\end{eqnarray*}
Equivalently, collecting  the bosonic one-forms $e^a$,
$B^{a_1a_2}$, $B^{a_1\cdots a_5}$ in (\ref{MCSigma(s)}) for $s=0$
in a symmetric spin-tensor one-form ${\cal E}^{\alpha\beta}$,
\begin{eqnarray}
 \label{cEff=def}
{\cal E}^{\alpha\beta}&=& {1\over 32}\left(e^a
\Gamma_{a}^{\alpha\beta} - {i\over 2}
B^{a_1a_2}\Gamma_{a_1a_2}{}^{\alpha\beta}  + {1\over 5!}
B^{a_1\ldots a_5} \Gamma_{a_1\ldots a_5}{}^{\alpha\beta} \right)\;
, \quad
\end{eqnarray}
the MC equations of $\tilde{\mathfrak{E}}(0)$ can be written as
\begin{eqnarray}\label{compacts0}
d{\cal E}^{\alpha\beta} = - i \psi^\alpha  \wedge \psi^\beta \; ,
\quad d\psi^\alpha =  0  \; , \quad d\eta^{\alpha} =  -64i \gamma_1
\, \psi^\beta \wedge {\cal E}_\beta{}^\alpha \; ; \quad
\end{eqnarray}
in the form given by eqs.~(\ref{osp-exp}) or (\ref{compacts0}) the
$Sp(32)$ automorphism symmetry of $\tilde{\mathfrak E}(0)$ becomes
manifest.

    For our purposes, the relevant features of the
 $\tilde{\mathfrak E}(s)$ superalgebras are the following:

\begin{enumerate}

\item For $s \neq 0$, the $\tilde {\mathfrak E}(s)$ may be considered
as deformations of $\tilde {\mathfrak E}(0)$.

     \item The automorphism group of $\tilde {\mathfrak E}(0)$ is
$Sp(32)$ while, for $s\neq0$, $\tilde {\mathfrak E}(s)$ has the
smaller $SO(1,10)$ group of automorphisms. Hence, the groups that
generalize the superPoincar\'e group
$\Sigma^{(11|32)}{\times\!\!\!\!\!\!\supset} SO(1,10)$, are given by
the following semidirect products

              \begin{itemize}

              \item $\tilde{\Sigma}(s)
{\times\!\!\!\!\!\!\supset} SO(1,10)$, $s \neq 0$,
 and

              \item  $\tilde{\Sigma}(0) {\times\!\!\!\!\!\!\supset}
SO(1,10) \approx Osp(1|32)(2,3,2)$,

              \item  $\tilde{\Sigma}(0) {\times\!\!\!\!\!\!\supset}
Sp(32) \approx Osp(1|32)(2,3)$,

              \end{itemize}

\end{enumerate}
where the last two right hand sides denote the appropriate {\it
expansions} of $OSp(1|32)$ (see later).


\subsection{ Trivialization of $\omega_4$}

To trivialize the CE four-cocycle $\omega_4 = - {1\over 4}
\Pi^\alpha \wedge \Pi^\beta \wedge \Pi^a \wedge \Pi^b
\Gamma_{ab}{}_{\alpha\beta}=d\omega_3$ (eq.(\ref{a4=0})), over the
$\tilde {\mathfrak E}(s)$ enlarged superalgebra one considers first
the most general ansatz  that expresses the three--form $A_3$ in
terms of combinations of wedge products of the one-forms $e^a$,
$\psi^\alpha$; $B^{a_1a_2}$, $B^{a_1 \ldots a_5}$, $\eta^\alpha$,
which are assumed to satisfy the MC equations
(\ref{MC1})--(\ref{deta=}). Using the same notation for MC forms and
fields here and below, we write
\begin{eqnarray}
\label{A3=Ans} 4A_3 &=& \lambda B^{ab} \wedge e_a \wedge e_b \; -
\alpha_1 B_{ab} \wedge B^b{}_c \wedge B^{ca} \nonumber \\
 &-& \alpha_2 B_{b_1a_1\ldots a_4} \wedge B^{b_1}{}_{b_2} \wedge
B^{b_2a_1\ldots a_4}
\nonumber \\
 &-& \alpha_3  \epsilon_{a_1\ldots a_5b_1\ldots b_5c} B^{a_1\ldots
a_5} \wedge B^{b_1\ldots b_5} \wedge e^c \nonumber \\
& - & \alpha_4 \epsilon_{a_1\ldots a_6b_1\ldots b_5} B^{a_1a_2
a_3}{}_{c_1c_2} \wedge B^{a_4a_5a_6c_1c_2} \wedge B^{b_1\ldots
b_5}
\nonumber \\
&-& 2i \psi^\beta \wedge \eta^\alpha \wedge \left( \beta_1 \, e^a
\Gamma_{a\alpha\beta} \right. \nonumber \\ && \left. \quad \qquad
\qquad -i \beta_2 \, B^{ab} \Gamma_{ab\; \alpha\beta} + \beta_3 \,
B^{abcde} \Gamma_{abcde\; \alpha\beta} \right) \; .
\end{eqnarray}
The problem is now to find the values of the constants $\alpha_1,
\ldots, \alpha_4$, $\beta_1, \ldots, \beta_3$ and $\lambda$, such
that eq.~(\ref{a4=}), $dA_3=a_4 =- {1\over 4} \psi^\alpha \wedge
\psi^\beta \wedge e^a \wedge e^b \Gamma_{ab}{}_{\alpha\beta}$, is
fulfilled. This produces a set of equations for the constants
$\alpha_1,\,\ldots,\, \beta_3$ and $\lambda$ that includes
$\delta$, $\gamma_1$ and $\gamma_2$ as parameters:

\begin{center}
\begin{tabular}{ll}
$\label{Eq1} \lambda - 2\delta\beta_1=1 \; ,$  & $\label{Eq2}
\lambda- 2\gamma_1 \beta_1 - 2\delta\beta_2
=0 \; , $ \\
$\label{Eq3} 3\alpha_1 + 8 \gamma_1\beta_2 =0 \; , $ &
$\label{Eq4} \alpha_2 - 10\gamma_1 \beta_3 - 10\gamma_2 \beta_2
=0 \; ,$ \\
$\label{Eq5} 5! \alpha_3 - \delta \beta_3 - \gamma_2 \beta_1 =0 \;
, $ & $\label{Eq6} \alpha_2 - 5!\,  10\gamma_2 \beta_3
=0 \; ,$ \\
$\label{Eq7} \alpha_3 - 2\gamma_2 \beta_3 =0 \; , $ &
$\label{Eq8} 3 \alpha_4 + 10\gamma_2 \beta_3 =0 \; . $
\end{tabular}
\end{center}
\begin{equation}
\end{equation}
This system has a nontrivial solution for
\begin{equation} \label{det}
\Delta = (2 \gamma_1-  \delta)^2 = 4s^2 \gamma_1^2 \not=0 \quad
\Longleftrightarrow \quad   s \neq 0   \; ,
\end{equation}
which in terms of the parameter $s$  reads \cite{CJS-D=11}
\begin{eqnarray}\label{sEq=g}
& \lambda =  {1 \over 5} \; \frac{s^2+2s+6}{s^2} \; , \; \nonumber
\\ &
 \beta_1 = -\frac{1}{10\gamma_1} \, \frac{2s-3}{s^2} \; , \;
\beta_2 = \frac{1}{20\gamma_1} \, \frac{s+3}{s^2} \; , \; \beta_3
= \frac{3}{10 \cdot 6! \gamma_1} \, \frac{s+6}{s^2}  \; ,
\nonumber \\
& \alpha_1 = -\frac{1}{15}\frac{2s+6}{s^2} \, ,
 \; \alpha_2 = \frac{1}{6!} \frac{{(s+6)}^2}{s^2} \,,
\alpha_3 = \frac{1}{5\cdot 5!}\alpha_2 \;, \
 \alpha_4 = -\frac{1}{9\cdot 5!} \alpha_2 \, ;
\end{eqnarray}
note that $\alpha_{2,3,4}\propto (s+6)$ and that all denominators
depend on $s^2$. Thus, $\omega_4$ can be trivialized ($\omega_4 =
d\tilde{\omega}_3$) over $\tilde{\mathfrak{E}}(s)$ when $s\not=0$;
the impossibility of doing it over $\tilde {\mathfrak E}(0)$ may be
related to the fact that precisely $\tilde {\mathfrak E}(0)$ has an
enhanced automorphism symmetry, $Sp(32)$. This implies that the
$A_3$ field can be considered as a composite of the one-form gauge
fields of any of the $\tilde{\Sigma}(s)$ with $s\not=0$, and that
$A_3$ is given by eq.~(\ref{A3=Ans}) for the values (\ref{sEq=g}) of
$\alpha_1, \ldots, \beta_3,\lambda$. Thus, the hidden gauge symmetry
of $D=11$ supergravity can be associated with any of the
$\tilde{\Sigma}(s){\times\!\!\!\!\!\!\supset}SO(1,10)$ supergroups.

The two particular solutions of D'Auria-Fr\'e for $A_3$ are
recovered by adjusting $\delta, \gamma_1$ in eq. (\ref{sEq=g}) so
that $\lambda=1$, which was the starting point of \cite{D'A+F}.
These correspond to $\tilde {\mathfrak E}(3/2)$, given by the
parameter values
\begin{eqnarray}\label{D'A+F1}  \quad  &
\delta = 5\gamma_1  \not=0 \quad ,
 \quad  \gamma_2=  {\gamma_1 \over 2\cdot 4!} \; ,
 \nonumber \\ & \lambda=1 \; , \; \beta_1=0 \; , \;\beta_2 = {1\over
10\gamma_1 } \; , \; \beta_3 = {1\over 6! \, \gamma_1} \; ,
\nonumber
\\ &  \alpha_1= - {4\over 15} \; , \; \alpha_2= {25\over 6!}  \; ,\;
\alpha_3= {1\over 6!\, 4! } \; , \; \alpha_4= - {1\over 54\, (4!)^2}
\; ,
\end{eqnarray}
and to $\tilde {\mathfrak E}(-1)$, for which
\begin{eqnarray}\label{D'A+F2}
 \quad &   \delta = 0 \; , \; \gamma_1 \not=0 \; ,
 \;  \gamma_2=  {\gamma_1 \over 3\cdot 4!} \; ,   \nonumber
\\  & \lambda=1 \; , \; \beta_1={1\over 2\gamma_1 } \;
, \; \beta_2 {1\over 10\gamma_1 } \; , \;  \beta_3 = {1\over 4^. 5!
\, \gamma_1}\; , \nonumber \\ & \alpha_1= - {4\over 15} \; ,\;
\alpha_2{25\over 6!}  \; ,\; \alpha_3= {1\over 6!\, 4! } \; ,\;
\alpha_4= - {1\over 54\, (4!)^2}\; . \quad
\end{eqnarray}

\subsection{The minimal solution
      ${\mathfrak E}_{min}$}

A specially simple trivialization of $\omega_4$ is achieved for
the superalgebra $\tilde {\mathfrak E}(-6)$, characterized by
\begin{eqnarray}
\label{S-6}
\tilde {\mathfrak E}(-6) &:& \quad \delta \neq 0 \; , \qquad
\delta= -10 \gamma_1 \; , \;   \gamma_2= 0 \; .
\end{eqnarray}
In $\tilde {\mathfrak E}(-6)$ the generator $Z_{a_1\ldots a_5}$ is
central (see eq.~(\ref{Sigma(s)})) and does not play any r\^ole in
the trivialization of the $\omega_4$ cocycle. Furthermore,
eqs.~(\ref{susyalg})--(\ref{idg}) allow us to use instead the
$\tilde {\mathfrak E}_{min}$ superalgebra whose central extension by
the generator $Z_{a_1\ldots a_5}$ gives $\tilde {\mathfrak E}(-6)$.
${\mathfrak E}_{min}$ is the $(66+64)$-dimensional superalgebra
$\tilde{\mathfrak E}^{(66|32+32)}$,
\begin{eqnarray}\label{susymin}
{\mathfrak E}_{min}:   & \{Q_\alpha,Q_\beta\}  =
\Gamma^a_{\alpha\beta} P_a + i\Gamma^{a_1a_2}_{\alpha\beta}
Z_{a_1a_2} \; ,
\\
 & [ P_a , Q_\alpha ]  =  -10 \gamma_1 \; \Gamma_{a \;
\alpha}{}^\beta Q^\prime_\beta \; , \quad [ Z_{a_1a_2} , Q_\alpha
] = i \gamma_1 \Gamma_{a_1a_2 \; \alpha}{}^\beta Q^\prime_\beta \;
, \quad
\end{eqnarray}
associated with the most economic
$\tilde{\Sigma}_{min}\equiv\Sigma^{(66|32+32)}$ generalized
supertranslation group that trivializes $\omega_4$.

 Using the values of eq.~(\ref{S-6}) in eq.~(\ref{sEq=g}) we get
\begin{eqnarray} \label{coeffminimal}
& \lambda=\frac{1}{6} \; , \; \beta_1= {1\over 4! \gamma_1 } \; ,\;
\beta_2 = -{1\over 2 \cdot 5! \gamma_1 } \; , \; \beta_3 = 0 \; ,
\nonumber \\ & \alpha_1= {1\over 90} \; , \; \alpha_2= 0  \; ,\;
\alpha_3= 0 \; , \; \alpha_4= 0 \; . \quad
\end{eqnarray}
Then, all the $B^{a_1 \ldots a_5}$ terms in $A_3$,
eq.~(\ref{A3=Ans}), are zero. This simplifies the expression for
$A_3$ drastically,
\begin{eqnarray}\label{A3minimal}
\label{A3=min} A_3 &=& {1\over 4!} B^{ab} \wedge e_a \wedge e_b \;
- {1\over 3^.5!} B_{ab} \wedge B^b{}_c \wedge B^{ca}
\nonumber \\
&- &
 {i \over 4^. 5! \, \gamma_1}  \psi^\beta \wedge \eta^\alpha
\wedge \left( 10 \, e^a \Gamma_{a\alpha\beta} + i \, B^{ab}
\Gamma_{ab\;\alpha\beta} \right) \; .
\end{eqnarray}
Thus, $\Sigma^{(66|32+32)}$ can be regarded as a {\it minimal}
underlying gauge supergroup of $D=11$ supergravity.


\section{Degrees of freedom in $D$=11 supergravity with a
composite $A_3$ and extra gauge gauge symmetries}

It remains to be checked that, as in the case of the BI fields of
the D-branes, the composite nature of $A_3$ does not change the
supergravity degrees of freedom. Let us first recall that, in
standard CJS supergravity, $e_\mu^a(x)$ has ${(D-2)(D-1) \over 2} -
1 = {D(D-3)\over2} =11^2 - 55_{Lorentz} - 2 \times 11_{Diff}= 44$
degrees of freedom; $\psi_\mu^\alpha (x)$ has $(9\times 32 - 32)
\times {1\over 2}= 128$; and $A_{\mu\nu\rho} (x)$ has  $\left(
\begin{matrix} 9 \cr 3\end{matrix} \right)= \left( \begin{matrix}
11 \cr 3\end{matrix}\right)- \left( \begin{matrix} 10 \cr
2\end{matrix}\right) - \left( \begin{matrix} 9 \cr 2\end{matrix}
\right)$ $= 84= 165_{\# \,\rm{of \, components}}$ --
$45_{(\#_{\rm{gauge \, symm.\;}} -\;\#_{\rm null \, vectors})}$ --
$36_{\#_{\rm{residual \,gauge\, symm.}}}$.

\medskip

Now, let us consider a composite $A_3$ in the CJS supergravity
action \cite{CJS} {\it i.e.}, by substituting
\begin{equation}
 A_3 = A_3 (B_1^{ab} \, , \, B_1^{a_1\ldots a_5} \, , \,
\eta_{1\alpha} \; ; \; e^a \, , \psi^\alpha) \; ,
\end{equation}
as given by eqs. (\ref{A3=Ans}) and (\ref{sEq=g}), for the
original three-form field in that action. Naively assuming
standard linearized equations and the usual `group theoretical'
gauge symmetry transformations for the new fields,
$$ \delta B_\mu^{ab} = \partial_\mu \alpha^{ab} + \ldots \, ;
\quad
 \delta B_\mu^{a_1\ldots a_5} = \partial_\mu \alpha^{a_1\ldots a_5} + \ldots \,;
 \quad
 \delta \eta_{\mu\alpha} = \partial_\mu \varepsilon^\prime_{\alpha} + \ldots
 \, ,
$$
the sum of the components of  $B_\mu^{ab}$ [$9 \times
\left(\begin{matrix} 11 \cr 2
\end{matrix} \right) = 495$], the components of
$B_\mu^{a_1\ldots a_5}$  [$9 \times \left(\begin{matrix} 11 \cr
5\end{matrix}\right)= 4158$] and those of $\eta_{\mu \alpha }$
[$128$, as for $\psi_\mu^\alpha$] would give a huge number of
`new' degrees of freedom for the gauge invariant theory with the
additional fields. Moreover, the bosonic and fermionic degrees of
freedom would not match.

However, the `new' fields $B_1^{ab} \, , \, B_1^{a_1\ldots a_5} \, ,
\, \eta_{1\alpha}$ enter in the CJS supergravity action only through
the composite $A_3(...)$ three form field and, as a result, the
theory possesses {\it extra gauge symmetries}. Clearly, these are
the transformations of the `new' fields that leave $A_3$ invariant,
\begin{eqnarray*}
\delta B_\mu^{ab} &=&
\partial_\mu \alpha^{ab} + \beta^{{}^{\left(
{}^{\fbox{}\fbox{}}_{\fbox{} }\right)}} {}_\mu^{ab} + \ldots \,
 \\
\delta B_\mu^{a_1\ldots a_5} &=& \partial_\mu \alpha^{a_1\ldots
a_5} + \beta_\mu^{a_1\ldots a_5}  + \ldots  \\
 \delta \eta_{\mu\alpha} &=& \partial_\mu \varepsilon^\prime_{\alpha} +
\beta_{\mu\alpha} +
 \ldots   \ .
\end{eqnarray*}
They reduce to $84$ the number of $B^{ab}_\mu$ degrees of freedom
and to zero those of the remaining new fields since,
diagrammatically (note that $\# {}^{\fbox{}\fbox{}}_{\fbox{}} =
D(D^2-1)/3 =440$)
\begin{eqnarray}\label{B=Yt}
B_{c\; ab} \; \sim \; {}^{\fbox{}} \otimes {}^{\fbox{}}_{\fbox{}}
= 11\times 55= 605= {}^{\fbox{}\fbox{}}_{\fbox{}} \oplus
\begin{matrix}\fbox{} \cr \vspace{-0.7cm} \cr \fbox{} \cr
\vspace{-0.68cm} \cr \fbox{}
\end{matrix} = 440+165\quad ,
\end{eqnarray}
and the equations of motion (which are the standard ones but with
a composite $A_3$), when linearized, affect only the antisymmetric
part $B_{[\mu\nu\rho]}$ of $B_\nu^{ab}$. In this way, the
antisymmetric 165-dimensional part simulates the fundamental
$A_3$; the mixed symmetry 440-dimensional part of $B_\nu^{ab}$ as
well as $B_\mu^{a_1\ldots a_5}$ and $\eta_{\mu\alpha}$ are pure
gauge and do not have independent equations of motion in the CJS
action with a composite $A_3$. Thus,
\begin{equation}
\nonumber \# \;{d.o.f.\;\rm{ with \; fundamental \;}  A_3} \, = \,
\#\; {d.o.f.\;\rm{ with \; composite \;} A_3 ,}
\end{equation}
as stated.


\section{The special element $\tilde {\mathfrak E}(0)$ as an algebra expansion}

We have seen that the superalgebra $\tilde {\mathfrak E}(0)$,
although it does not trivialize $\omega_4$, may be considered as a
`parent' superalgebra for the hidden symmetries of $D=11$
supergravity in the sense that it gives rise to the family $\tilde
{\mathfrak E}(s)$ of superalgebras that do trivialize the standard
supersymmetry algebra ${\mathfrak E}$ four-cocyle. All the
corresponding $\tilde{\Sigma}(s)$ enlarged superspace groups, $s\neq
0$, may be considered as deformations of $\tilde{\Sigma}(0)$. We
shall now characterize the parent algebra $\tilde {\mathfrak E}(0)$
in terms of Lie algebra expansions. With this aim, we first review
briefly the expansion method \cite{AIPV02,H01} for the case which is
of special interest here.

\subsection{The algebra expansion method }

Let $G$ be a Lie group, of local coordinates $g^i$, $\mathcal{G}$
its Lie algebra and  ${\mathcal G}^*$ its dual coalgebra. Let
$\mathcal{G}$ admit, say, the splitting $\mathcal{G}=V_0\oplus V_1
\oplus V_2$, where $V_0$, $V_2$ ($V_1$), are even (odd) subspaces of
dimensions $\textrm{dim}\, V_p$, $p=0,1,2$. Further, let $V_0$ be a
subalgebra of $\mathcal{G}$ and $[V_1 , V_1] \subset V_0 \oplus
V_2$, $[V_2, V_2] \subset V_0 \oplus V_2$ (details for the general
theory are given in \cite{AIPV02}). Then, the rescaling of the group
parameters $g^{i_p} \rightarrow \lambda^p g^{i_p}$, $i_p=1,\ldots,
\textrm{dim}\, V_p$ , allows us to expand the one-forms
$\omega^{i_p}(\lambda,g)$, obtained from the algebra MC forms
$\omega^{i_p}(g)$ that define a basis of the dual subspaces
${V_p}^*$, as a series in $\lambda$,
\begin{equation} \label{expWW}
\omega^{i_p}(\lambda) = \lambda^p \omega^{i_p , p} + \lambda^{p+2}
\omega^{i_p , p+2} + \lambda^{p+4} \omega^{i_p , p+4} + \ldots \,
=\sum_{\alpha_p} \lambda^{\alpha_p} \omega^{i_p,\alpha_p} \;\;
(p=0,1,2)\;.
\end{equation}
The different powers of lambda  are a consequence of the above
assumptions on the subspaces $V_p$, and follow from the fact that
the canonical form $\theta(g)$ on a Lie group $G$ is given by
$\theta(g)=g^{-1}dg=\omega^iX_i$, where $g={\textrm exp}\;{g^i X_i}$
and $X_i$ are the generators of the algebra ${\cal G}$ of $G$. The
insertion of these series expansions into the MC equations of the
original algebra $\cal G$,
\begin{eqnarray} \label{eq:MC}
& d\omega^{i_p}= -\frac{1}{2} c_{j_q k_s}^{i_p} \, \omega^{j_q}
\wedge \omega^{k_s} \nonumber
\\ & (p,q,s=0,1,2 \; ; \; i_{p,q,s}=1,2,\ldots,
\textrm{dim} \, V_{p,q,s}) \; ,
\end{eqnarray}
produces, identifying the terms with the same order in $\lambda$,
the following set of equations
\begin{eqnarray}
\label{eq:MCap}
& d\omega^{i_p,
\alpha_p}= -\frac{1}{2} C_{j_q,\beta_q \;
k_{s},\gamma_s}^{i_p,\alpha_p}\; \omega^{j_q, \beta_q}
\wedge \omega^{k_s, \gamma_s} \;, \nonumber \\
& C_{j_q,\beta_q \; k_s,\gamma_s}^{i_p,\alpha_p}= \left\{
\begin{array}{lll} 0, &
\mathrm{if} \ \beta_q + \gamma_s \neq \alpha_p  \\
c_{j_q k_s}^{i_p}, & \mathrm{if} \ \beta_q + \gamma_s = \alpha_p
\end{array} \right. \nonumber \\
& (\alpha_p,\beta_p,\gamma_p=p,p+2, \ldots  ) \quad.
\end{eqnarray}

    The question now is how to retain consistently a number of
$\omega^{i_p, \alpha_p}$ so that the equations above correspond to
the MC equations of a new, by construction expanded, algebra.
Cutting the expansions of the $\omega^{i_p}(\lambda)$ at certain
orders $\alpha_p=N_p,\; p=0,1,2$, one finds that
eqs.~(\ref{eq:MCap}) for $\alpha_p=p,\ldots,N_p$ will provide the
MC equations of a new finite-dimensional Lie algebra provided the
chosen orders satisfy the conditions
\begin{equation}
\label{conWW} N_0  =  N_1+1 = N_2 \qquad \mathrm{or} \qquad N_0  =
N_1-1 = N_2 \qquad \mathrm{or} \qquad  N_0  =  N_1-1= N_2-2 \; .
\end{equation}
These conditions guarantee that for the selected set of
$\omega^{i_p, \alpha_p}$'s, eqs. (\ref{eq:MCap}) do not include
any $\omega^{i_p, \alpha_p}$ outside this set and that,
accordingly, define new algebras \cite{H01,AIPV02} by becoming
their MC equations. These algebras, denoted
$\mathcal{G}(N_0,N_1,N_2)$ in obvious notation, are called {\it
expansions} of $\mathcal{G}$; in general, their dimension is
larger than that of the original algebra $\mathcal{G}$. They also
include, as a particular case and for a specific value of
($N_0,N_1,N_2$), the generalized Wigner-\.In\"on\"u contractions
\cite{AIPV02}, in which case the dimension does not change.

The dimension of the expanded
 $\mathcal{G}(N_0,N_1,N_2)$ algebras is given by

\begin{eqnarray} \label{eq:dimsuper}
\textrm{dim}\,\mathcal{G}(N_0,N_1,N_2) &=&
\left[(N_0+2)/{2}\right]\textrm{dim}V_0 +
\left[(N_1+1)/{2}\right]\textrm{dim}V_1  \nonumber
\\&& +\left[N_2/{2}\right]\textrm{dim}V_2 \quad .
\end{eqnarray}


\subsection{$\tilde{\Sigma}(0){\times\!\!\!\!\!\!\supset}SO(1,10)$
as the expansion $OSp(1|32)(2,3,2)$ }

Let us now consider the orthosymplectic algebra $osp(1|32)$, of
dimension 560, defined by the MC equations
\begin{eqnarray}
     d\rho^{\alpha\beta} &=& -i\rho^{\alpha\gamma}\wedge
\rho_\gamma{}^\beta-i\nu^\alpha\wedge \nu^\beta \; , \nonumber\\
d\nu^\alpha &=& -i\nu^\beta \wedge \rho_\beta{}^\alpha\; , \quad
\alpha,\beta=1,\ldots,32 \quad , \label{MCosp132}
\end{eqnarray}
where the 528 $\rho^{\alpha\beta}$ are the bosonic and the 32
$\nu^\alpha$ the fermionic MC one-forms.

 The decomposition of $\rho_{\alpha\beta}$ as
\begin{equation}
\rho_{\alpha\beta}= \frac1{32} \left(  \rho^a \Gamma_a
-\frac{i}{2} \rho^{ab} \Gamma_{ab}+ \frac{1}{5!} \rho^{a_1\dots
a_5} \Gamma_{a_1\dots a_5}\right)_{\alpha\beta} \; , \;\;
a,b=0,1,\ldots,10 \; .\label{generalrho}
\end{equation}
allows us to consider the splitting $osp(1|32)=V_0\oplus V_1\oplus
V_2$, where
\begin{eqnarray}
\begin{array}{lc}
V_0^* \quad \textrm{is generated by} \quad  \rho^{ab} \quad & (55)
\quad ,
\nonumber \\
V_1^* \quad \textrm{by} \quad  \nu^\alpha
& (32) \quad, \nonumber \\
V_2^* \quad \textrm{by} \quad \rho^a \; \textrm{and} \;
\rho^{a_1\dots a_5} & (11+462) \quad .
\end{array}
\end{eqnarray}

The various forms then expand as
\begin{eqnarray}
V_0^*:\quad &&\rho^{ab}=\rho^{ab,0}+\lambda^2
\rho^{ab,2}+\cdots;\quad V_1^*:\quad
\nu^\alpha=\lambda\nu^{\alpha,1} +
\lambda^3\nu^{\alpha,3}+\cdots; \nonumber\\
V_2^*:\quad &&\rho^a =\lambda^2 \rho^{a,2}+\cdots ,\qquad
\rho^{a_1\dots a_5}=\lambda^2 \rho^{a_1\dots a_5,2}+\cdots\; .
\label{fullMexp}
\end{eqnarray}
Inserting the series into the MC equations and choosing $N_0=2$,
$N_1=3$, $N_2=2$ the MC equations of the expansion
$osp(1|32)(2,3,2)$ are obtained: {\small \begin{eqnarray}
\label{ospmaurerd}
      d\rho^{ab,0} &=& -\frac{1}{16}
 \rho^{ac,0}\wedge {\rho_c}^{b\,,0}  \nonumber\\
      d\rho^{a\,,2} &=& -\frac{1}{16} \rho^{b,2}\wedge
{\rho_b}^{a,0} -i \nu^{\alpha,1}
\wedge \nu^{\beta,1} \Gamma^{a}_{\alpha\beta} \nonumber\\
    d\rho^{ab,2} &=&  -\frac{1}{16} \left(  \rho^{ac,0}\wedge
{\rho_c}^{b,2} + \rho^{ac,2}\wedge {\rho_c}^{b,0} \right) -
\nu^{\alpha,1}
\wedge \nu^{\beta,1} \Gamma^{ab}_{\alpha\beta} \nonumber\\
      d\rho^{a_1\dots a_5}{}^{,2} &=& \frac{5}{16}
\rho^{b[a_1\dots a_4|\,,2} \wedge \rho_b{}^{|a_5],0}\nonumber - i
\nu^{\alpha,1} \wedge \nu^{\beta,1} \Gamma^{a_1\dots
a_5}_{\alpha\beta}
\nonumber\\
     d\nu^{\alpha,1} &=& -\frac{1}{64} {\nu}^{\beta,1} \wedge
\rho^{ab,0}{\Gamma_{ab}}_\beta{}^\alpha
\nonumber\\
     d\nu^{\alpha,3} &=& -\frac{1}{64} {\nu}^{\beta,3} \wedge
\rho^{ab,0}{\Gamma_{ab}}_\beta{}^\alpha \\&& - \frac1{2}
\nu^{\beta,1} \wedge {\left(i\rho^{a,2}\Gamma_a +
\frac{1}{2}\rho^{ab,2}\Gamma_{ab} + \frac{i}{5!}\rho^{a_1\dots
a_5,2} \Gamma^{a_1\dots a_5}\right)_\beta}^\alpha  \ . \nonumber
\end{eqnarray}}

With the identifications
\begin{eqnarray}\label{identifications}
\rho^{ab,0} &=& -16 \omega^{ab} \quad,\quad \rho^{a,2}=e^a \quad,
\quad \rho^{ab,2}=B^{ab},\nonumber \\
\rho^{a_1 \cdots a_5,2}&=&B^{a_1\cdots a_5} \quad, \quad
\nu^{\alpha,1}=\psi^\alpha \quad, \quad \nu^{\alpha,3} =
\eta^\alpha/64\gamma_1 \quad,
\end{eqnarray}
and omitting the Lorentz generators $\omega^{ab}$ to simplify, these
equations read
\begin{eqnarray}
\label{Sigma(s=0)}
de^a &=& - i\psi^{\alpha} \wedge \psi^{\beta}
\Gamma^a_{\alpha\beta} \; , \nonumber \\
 dB^{a_1a_2} &=& - \psi^\alpha \wedge \psi^\beta \,
\Gamma^{ab}_{\alpha\beta} \; , \nonumber \\
  dB^{a_1\ldots a_5} &=& - i \psi^\alpha \wedge
\psi^\beta \, \Gamma^{a_1\ldots a_5}_{\alpha\beta} \; , \nonumber  \\
d\psi^\alpha &=& 0 \; , \;    \\
d\eta^{\alpha} &=& -2\gamma_1 \cdot \psi^\beta \wedge {\left( i \,
e^a \Gamma_{a}  + \frac{1}{2} \, B^{ab} \Gamma_{ab} + \frac{i}{5!}
\, B^{a_1\ldots a_5} \Gamma_{a_1\ldots a_5} \right)_\beta}^\alpha
\quad ; \nonumber
\end{eqnarray}
the inclusion of $\omega^{ab}$ produces the MC equations of the
$\tilde{\mathfrak{E}}(0){\times\!\!\!\!\!\!\supset} so(1,10)$
algebra. Also, one may check that
\begin{eqnarray}
&&\mathrm{dim}\,OSp(1|32)(2,3,2)=2\cdot55+2\cdot
32+1\cdot473=647=\nonumber\\
&&=592+55=\mathrm{dim}\,(\tilde\Sigma(0){\times\!\!\!\!\!\!\supset}
SO(1,10)) \quad .
\end{eqnarray}


\subsection{ $\tilde{\Sigma}(0) {\times\!\!\!\!\!\!\supset} Sp(32)$
as the expansion $OSp(1|32)(2,3)$ }

Let us now see that the full $\tilde{\mathfrak{E}}(0)
{+\!\!\!\!\!\!\supset} sp(32)$ is also an expansion of
$osp(1|32)$. Let us consider now the splitting
$osp(1|32)=V_0\oplus V_1$ where $V_0$ is the bosonic subalgebra,
generated by $\rho^{\alpha\beta}$, and $V_1$ the fermionic part,
generated by $\nu^\alpha$. Choosing $N_0=2$ and $N_1=3$ we obtain
the expansion $osp(1|32)(2,3)$ determined by the MC equations of
the one-forms $\rho^{\alpha\beta,0}$, $\rho^{\alpha\beta,2}$,
$\nu^{\alpha,1}$, $\nu^{\alpha,3}$:
\begin{eqnarray}\label{MC-sp}
&& d\rho^{\alpha\beta,0}= -i\rho^{\alpha\gamma,0}\wedge
\rho_\gamma{}^{\beta ,0} \nonumber\\
&& d\nu^{\alpha,1} = - i\nu^{\beta,1} \wedge \rho_\beta{}^{\alpha
, 0}
 \nonumber\\
&& d\rho^{\alpha\beta,2}= -i \left( \rho^{\alpha\gamma,0}\wedge
\rho_\gamma{}^{\beta ,  2} + \rho^{\alpha\gamma,2}\wedge
\rho_\gamma{}^{\beta ,  0} \right) -i\nu^{\alpha,1}\wedge \nu^{\beta,1} \nonumber\\
&& d\nu^{\alpha,3} = -i \nu^{\beta,3} \wedge \rho_\beta{}^{\alpha
,  0} - i\nu^{\beta,1} \wedge \rho_\beta{}^{\alpha,2}\;.
\end{eqnarray}
Identifying $\rho^{\alpha\beta,0}$ in eqs. (\ref{MC-sp}) with the
$sp(32)$ connection $\Omega^{\alpha\beta}$, eqs, (\ref{MC-sp})
coincide with those of
$\tilde{\mathfrak{E}}(0){+\!\!\!\!\!\!\supset} sp(32)$
[eqs.~(\ref{compacts0})], with the identifications
$\rho^{\alpha\beta,2}={\cal E}^{\alpha\beta}$,
$\nu^{\alpha,1}=\psi^\alpha$ and
$\nu^{\alpha,3}=\eta^\alpha/64\gamma_1$. One can also make a
dimensions check:
\begin{eqnarray}
&&\textrm{dim}(\tilde{\mathfrak{E}}(0) {+\!\!\!\!\!\!\supset}
sp(32))=592 \;(528+64)+528=1120=\nonumber\\&&=2\cdot528+2\cdot32
=\textrm{dim}(\, osp(1|32)(2,3))
\end{eqnarray}
when $N_0=2$, $N_1=3$ in eq.~(\ref{eq:dimsuper}).


\section{ Conclusions}

We have given some reasons in favour of a geometrical {\it
enlarged superspace coordinates/fields correspondence}, both for
branes, in which case the correspondence is between the extended
superspace coordinates and {\it worldvolume} fields, and for
$D=11$ CJS supergravity, where the fields are {\it spacetime}
fields.

In the case of {\it branes}, the new enlarged superspace algebras
appear as the result of wishing to have manifestly invariant WZ
terms or an (enlarged) superspace origin for all the fields of the
theory, including the otherwise `intrinsically' worldvolume fields
of the $D$-branes (Born-Infeld fields) and of the M5 brane. The CE
cohomology arguments that lead to the WZ terms for the scalar
$p$-branes also allow us to caracterize the D-branes as well as
the WZ term of the M5-brane. Their actions (apart from the
auxiliary field in the M5-brane case) do not contain fields
directly defined on the worldvolume\footnote{Although in the
(either rigid or non-flat) $D=11$ covariant M5-brane action
\cite{BPSTV95} the Pasti-Sorokin-Tonin (PST) scalar $a(\xi)$ field
\cite{Pa-So-To-97} is a {\it worldvolume} field, it is an {\it
auxiliary} one. Further, it was shown in \cite{Ba-Be-So-97} that,
when the M5-brane interacts with dynamical supergravity in a
duality symmetric formulation, the r\^ole of the M5-brane
auxiliary PST scalar $a(\xi)$ is played by the pull-back
$a(x(\xi))$ to ${\cal W}$ of the {\it spacetime} supergravity PST
scalar $a(x)$ and is a kind of background field.}; all worldvolume
fields are associated to variables of certain enlarged superspaces
$\tilde{\Sigma}$. Further, the number of degrees of freedom and
the dynamical contents of the E-L equations remain the same once
the substitution is made \cite{A-I-bra-00}.

   The fields/coordinates correspondence for $D=11$ {\it CJS
supergravity} has also to do with trivializing non-trivial CE
cocycles. Trivializing the supersymmetry algebra
$\mathfrak{E}^{(11|32)})$ CE four--cocycle $\omega_4$ amounts to
finding a composite structure for the three--form field $A_3$ of the
standard Cremmer--Julia--Scherk supergravity in terms of one--form
gauge fields of $\tilde{\Sigma}(s)$, $A_3 = A_3(e^a \, , \,
\psi^\alpha\,$; $B^{a_1a_2}$, $B^{a_1 \ldots a_5}, \eta^\alpha \,)$.
The trivialization of the CE four-cocycle $\omega_4$ may be
achieved, for $s \neq 0$, on the one-parametric family of
superalgebras $\tilde{\mathfrak{E}}(s)$. These are central
extensions of the M-algebra (generated, ignoring the Lorentz part,
by $P_a$, $Q_\alpha$, $Z_{ab}$, $Z_{a_1\ldots a_5}$) by an
additional fermionic central generator $Q^\prime_\alpha$. Then,
$\omega_4 = d\tilde{\omega}_3(\tilde{Z})$, $\tilde{Z} \in
\tilde{\Sigma}$. The Maurer-Cartan forms of
$\tilde{\mathfrak{E}}(s)$  can be replaced by {\it soft} one-forms
obeying a free differential algebra with curvatures, and thus {\it
one may treat the standard CJS D=11 supergravity as a gauge FDA of
the $\tilde{\Sigma}(s)$ supergroup for any $s\neq 0$}.
 This fact was known before for the two superalgebras
\cite{D'A+F} that here correspond to $\tilde{\mathfrak{E}}(3/2)$
and $\tilde{\mathfrak{E}}(-1)$. The novelty of the present results
is that, for $s\neq0$, any of the $\tilde{\Sigma}(s)
{\times\!\!\!\!\!\!\supset} SO(1,10)$ supergroups may be equally
treated as an underlying gauge supergroup of the $D=11$
supergravity.

  There is a special element in the $\tilde{\mathfrak{E}}(s\neq 0)$
family of trivializations, $\tilde{\mathfrak{E}}(-6)$, for which
the $Z_{a_1\ldots a_5}$ generator is central. In this case, the
expression for $A_3$ is particularly simple: it does not involve
the one-form $B^{a_1\ldots a_5}$, and $\tilde{\mathfrak{E}}(-6)$
may be reduced to ${\mathfrak E}_{min}$. Thus, the smaller
$\tilde{\Sigma}_{min}= \tilde{\Sigma}^{(66|32+32)}$ associated
with ${\mathfrak E}_{min}$ may be considered as the minimal
underlying gauge supergroup of $D=11$ CJS supergravity. All other
representatives of the family $\tilde{\mathfrak{E}}(s)$ are
equivalent, although they are not isomorphic. Their significance
might be related to the fact that the field $B^{a_1\ldots a_5}$ is
also needed for a coupling to BPS preons \cite{BPS01,BPS04}, the
hypothetical basic constituents of M-theory. The presence of a
full family of superalgebras $\tilde{\mathfrak{E}}(s\neq 0)$
--rather than a unique one-- trivializing the standard
$\mathfrak{E}^{(11|32)}$ algebra four--cocycle $\omega_4$,
suggests that the obtained underlying gauge symmetries of $D=11$
supergravity may be incomplete (this is almost certainly the case
if one considers the symmetries of M-theory).

The singularity of the $\tilde{\mathfrak{E}}(0)$ case looks a
reasonable one; the $\tilde{\Sigma}(0)$ supergroup is special
because it possesses an enhanced automorphism symmetry, $Sp(32)$.
The full ${\tilde{\Sigma}(0) {\times\!\!\!\!\!\!\supset} Sp(32)}$,
that replaces the $D=11$ superPoincar\'e group, is given by the
expansion $OSp(1|32)(2,3)$ of $OSp(1|32)$. All other members of
the $\tilde{\Sigma}(s\neq 0)$ family have the smaller $SO(1,10)$
automorphism symmetry and are deformations of the $s=0$ element.
Thus, we may conclude that {\it the underlying gauge group of
$D=11$ supergravity is determined by any element
${\tilde{\Sigma}(s\neq 0){\times\!\!\!\!\!\!\supset} SO(1,10)}$,
of a one-parametric familiy of nontrivial deformations of
${\tilde{\Sigma}(0){\times\!\!\!\!\!\!\supset} SO(1,10)} \approx
OSp(1|32)(2,3,2) \subset
{\tilde{\Sigma}(0){\times\!\!\!\!\!\!\supset} Sp(32)} $ }.
Furthermore, we see that the number of the extended superspace
coordinates are in one-to-one correspondence with the gauge fields
entering the theory, and that the additional degrees of freedom
may be removed by a gauge transformation. Thus, this may be
considered as another example of the conjectured extended
superspaces coordinates/fields correspondence principle in which
the fields are spacetime fields.

Finally it is known that, unlike its lower dimensional versions,
CJS supergravity forbids a cosmological term extension. The reason
is cohomological and can be traced to an obstruction produced by
the $A_3$ three-form field \cite{BaDeHeSe97}. It is natural to ask
wether this obstruction remains when $A_3$ is becomes a composite
field.

\noindent {\bf Acknowledgments}. This contribution is based on
research mostly done in collaboration with I. Bandos, J.M.
Izquierdo, M. Pic\'on and O. Varela, refs.
\cite{A-I-bra-00,AIPV02,CJS-D=11}, which is acknowledged with
pleasure. This work has been partially supported by the research
grant BFM-2002-03681 from the Spanish Ministerio de Educaci\'on y
Ciencia and from EU FEDER funds, the Generalitat Valenciana
(Grupos 03/124), and by the EU network MRTN--CT--2004--005104
(`Forces Universe').


\end{document}